\documentclass[draft,pra,aps]{revtex4}

\begin{document}

\title{Generalized RECPs accounting for Breit effects: uranium, plutonium 
 and superheavy elements 112, 113, 114}

\author{N.S. Mosyagin}\email{mosyagin@pnpi.spb.ru}
\author{A.N. Petrov}
\author{A.V. Titov}
\affiliation{Petersburg Nuclear Physics Institute, 
             Gatchina, St.-Petersburg district 188300, Russia}
\author{I.I. Tupitsyn}
\affiliation{Physics Department, St.-Petersburg State University, 
Starii Petergoff, St.-Petersburg 198904, Russia}

\begin{abstract}
 The Generalized Relativistic Effective Core Potential (GRECP) method is
 described which allows one to simulate Breit interaction and finite nuclear
 models by an economic way and with high accuracy.  The corresponding GRECPs
 for the uranium, plutonium, eka-mercury (E112), eka-thallium (E113) and
 eka-lead (E114) atoms are generated.  The accuracy of these GRECPs and of the
 RECPs of other groups is estimated in atomic numerical SCF calculations with
 Coulomb two-electron interactions and point nucleus as compared to the
 corresponding all-electron Hartree-Fock-Dirac-Breit calculations with the
 Fermi nuclear charge distribution.
  Different nuclear models and contributions of the Breit interaction between
  different shells are studied employing all-electron four-component methods.
\end{abstract}

\maketitle

\section*{Introduction}

 Investigation of physical and chemical properties of recently synthesized
 relatively long-living isotopes of superheavy elements (SHEs)
 with the nuclear charges $Z$=105 to 116
 \cite{Hoffman:00, Oganessian:99, Oganessian:01, Schadel:03b} 
 and their compounds is of fundamental importance for
  science.
 Their experimental lifetimes reach several hours now and the nuclei near the
 top of the ``island of stability'' are predicted to exist for many years.  The
 experimental study of SHE properties is very difficult because of their
 extremely small quantities, only single atoms are available for research.
 Accurate calculations for SHEs and their compounds are needed in order to
 better understand their physical and chemical properties that
  often differ
 from those of the lighter homologs in the chemical groups due to very strong
 relativistic effects on their electronic shells.  Besides, for elements
 decaying by spontaneous fission, the chemical identification is the only way
 to prove their $Z$ number.

 Experimental investigations of spectroscopic and other physical-chemical
 properties of actinides are severely hampered by their radioactive decay and
 radiation which lead to chemical modifications of the systems under study.
 The diversity of properties of lanthanide and actinide compounds is unique due
 to the multitude of their valency forms (which can vary over a wide range) and
  because of
 particular importance of relativistic effects.
 They are, therefore, of great interest both for fundamental research and for
 development of new technologies and materials. The most important practical
 problems involve storage and processing of radioactive waste and nuclear fuel,
 as well as pollution of the environment by radioactive waste, where most of
 the decayed elements are actinides.

 From the formal point of view, four-component correlation calculations
 \cite{Hirao:04,Schwerdtfeger:04aa} based on Dirac-Coulomb-Breit (DCB)
 Hamiltonian (see \cite{Mohr:97, Grant:00b, Reiher:00, Shabaev:02a,
 Labzowsky:02b} and references) can provide a very high accuracy of physical
 and chemical properties for molecules containing heavy atoms. However, such
 calculations were not widely used for such systems
  during last decade
 because of the following theoretical and technical complications
 \cite{Visscher:96}:

\begin{itemize}
 \item[{\bf -}]
         too many electrons are treated explicitly in heavy-atom sys\-tems and
         too large number of Gaussians is required for accurate description of
         the large number of oscillations, which valence spinors have in
         heavy atoms;

 \item[{\bf -}]
         the necessity to work with four-component Dirac
         spinors leads to serious complication of calculations as compared to
         the nonrelativistic case:

 \begin{itemize}
  \item[(a)]
         the number of kinetically-balanced two-component (``$2c$'')
         uncontracted Gaussian basis spinors for the {\it S}mall components,
         $N_S^{2c}$, can be estimated as $2N_L^{2c}$, where $N_L^{2c}$ is the
         number of basis spinors for {\it L}arge components; so the total
         number of uncontracted Gaussian basis spinors in the relativistic
         four-component (``$4c$'') calculations $N_{bas}^{4c} \sim 3 N_L^{2c}$
         and the number of two-electron integrals as
         \cite{Visscher:96}
         $$
            N_{2eInt}^{4c} \sim (1{+}2{\cdot}2^2{+}2^4) N_{2eInt}^{2c}
                           \equiv 25{\cdot}N_{2eInt}^{2c}\ ;
         $$
         Note, however, that
         the situation is seriously improved here during last years, see
         \cite{Dyall:02a, Visscher:02aa, Hirao:04, Schwerdtfeger:04aa}.
  \item[(b)]
         the number of basis $2c$-spinors, $N_{bas}^{2c}$, is twice more than
         the number of nonrelativistic basis one-component
         (``$1c$'') orbitals, $N_{bas}^{1c}$, therefore
         $$
            N_{2eInt}^{2c} \sim 2^4/2{\cdot}N_{2eInt}^{1c}
                           \equiv 8{\cdot}N_{2eInt}^{1c}\ ,
         $$
         The minimal number of two-electron integrals in the {spin-orbit} basis
         set, which are required to be saved
         coincides, obviously, with $N_{2eInt}^{1c}$.
 \end{itemize}

\end{itemize}

 The Relativistic Effective Core Potential (RECP) method is most widely used in
 calculations on molecules containing heavy atoms 
 \cite{Ermler:88, Titov:05b} because it
 reduces drastically the computational cost at the integral generation, SCF and
 integral transformation stages.  In our papers \cite{Tupitsyn:95, Mosyagin:97,
 Titov:99}, the conventional radially-local (semi-local) form of the RECP
 operator (used by many groups up to now but suggested and first applied about
 40 years ago \cite{Phillips:59, Abarenkov:65, Heine:64}) was shown to be
 limited by accuracy and some nonlocal corrections to the RECP operator were
 suggested \cite{Tupitsyn:95, Titov:95, Titov:99, Titov:00a}, which have already
 allowed us to improve significantly the RECP accuracy
 \cite{Mosyagin:97,Titov:99,Mosyagin:00,Isaev:00}.

 It is known that the Breit interaction can give contributions in excess of one
 thousand wave numbers even to energies of transitions between lowest-lying
 states of very heavy elements (see, e.g., tables \ref{U_conf} and
 \ref{Pu_conf}).  It is also clear that the point nuclear model becomes less
 appropriate when the nuclear charge is increased.  Therefore, the RECPs
 designed for accurate calculations of actinide and SHE compounds should allow
 one to take into account the Breit interaction and the finite size of nuclei.
 The most economic way is to incorporate the corresponding contributions into
 the RECP operator.

\section{Generalized RECP method}

 In a series of papers (see \cite{Tupitsyn:95, Mosyagin:97, Titov:99, Titov:00a,
 Titov:02Dism} and references), we
 introduced and developed the Generalized RECP (GRECP) method.  Its main
 features are:

\begin{itemize}
\item
 The inner core (IC), outer core (OC) and valence (V) electrons are first
 treated employing different approximations for each (including relaxation of
 the IC shells which are explicitly excluded from GRECP calculations).

\item
 GRECP involves both radially-local, separable and Huzinaga-type potentials as
 its components and particular cases.

\item
 The GRECP operator includes terms of other types for economical treatment of
 transition metals, lanthanides and actinides (see sections
 \ref{sUSfC}--\ref{sTScor}).

\item
   The outer core pseudospinors (nodeless) together with valence pseudospinors
   (nodal) are used for constructing the GRECP components \cite{Titov:91}.

\item
   Quantum electrodynamics effects (see \cite{Petrov:04b} and section
   \ref{sDCBsh}),
   arbitrary nuclear models, and correlation with IC shells \cite{Mosyagin:04a}
   can be efficiently treated within GRECPs.
\end{itemize}
 The GRECP method is described in detail in the above papers and we only add
 here that
  it
 allows one to avoid the complications of the four-component calculations
 described in the introduction
  (see also \cite{Mosyagin:04a}) and to attain very high accuracy, limited in
  practice by possibilities of the correlation methods,
 while requiring moderate computational efforts when the IC, OC and V subspaces
 are appropriately chosen.

 The contributions of different nuclear models which are described by local
 potentials can be easily taken into account in the framework of the (G)RECP
 method.  The situation is more complicated in the case of the Breit
 interaction because it is represented by a two-electron operator.  General
 justification of the possibility to simulate the Breit effects by means of an
 one-electron (G)RECP operator with good accuracy and the scheme of such GRECP
 generation are presented in \cite{Petrov:04b} (see also section \ref{sDCBsh}).
 This scheme is applied in the present work to generate GRECPs for the uranium,
 plutonium, eka-mercury (E112), eka-thallium (E113) and eka-lead (E114) atoms.
 The 32, 34, 20, 21 and 22 electrons are explicitly treated in calculations
 with these GRECPs, correspondingly.  Moreover, the 52 electron GRECP
 (52e-GRECP) version for E112 was also generated. The conventional Coulomb
 operator for two-electron interactions and the point nuclear model should be
 used in these GRECP calculations.  However, they will account for 
  the Fermi nuclear charge model that is close to the experimental
  distribution.
 Moreover, the Breit interactions of the electrons from the state used for the
 GRECP generation with the electrons explicitly treated in the subsequent
 calculations are simulated by the GRECP (in some sense, the Breit interaction
 is ``frozen'' here).

\subsection{Self-consistent GRECP version for $d$- and $f$-elements}
\label{sUSfC}

 The Self-Consistent (SfC) (G)RECP version \cite{Titov:95, Titov:99, Titov:00a,
 Titov:02Dism} allows one to minimize errors for energies of transitions with
 the change of the occupation numbers for the OuterMost Core (OMC) shells
 without extension of space of explicitly treated electrons.  It allows one to
 take account of relaxation of those core shells, which are explicitly excluded
 from the GRECP calculations, thus going beyond the frozen core approximation.
 This method is most optimal for studying compounds of transition metals,
 lanthanides, and actinides.  Features of constructing the self-consistent
 GRECP are:
\begin{enumerate}
\item   The all-electron HFDB calculations of two generator states with
        different occupation numbers $N_1$ and $N_2$ of the 
        OMC $d$ or $f$ shell are carried out for an $d$- or $f$-element.
\item   The GRECP versions with separable correction
        ${\bf U}^{N_1}$ and ${\bf U}^{N_2}$ are constructed for
        these generator states employing the standard scheme 
	\cite{Tupitsyn:95,Mosyagin:97,Titov:99,Titov:00a}.
        The GRECP operator with the separable correction has the form

\begin{eqnarray}
 \label{UGRECP}
  {\bf U}^{N_i}  &=&  E_{\rm core}^{N_i} + U_{n_vLJ}^{N_i}(r)
                 +  \sum_{l=0}^L \sum_{j=|l-1/2|}^{l+1/2}
		   \Bigl\{\bigl[U_{n_vlj}^{N_i}(r)
		    -  U_{n_vLJ}^{N_i}(r)\bigr] {\bf P}_{lj}   \nonumber\\
                &+&   \sum_{n_c}
                   \bigl[U_{n_clj}^{N_i}(r)
		   -  U_{n_vlj}^{N_i}(r)\bigr]
                   \widetilde{\bf P}_{n_clj}^{N_i}
		   +   \sum_{n_c}  \widetilde{\bf P}_{n_clj}^{N_i}
                   \bigl[U_{n_clj}^{N_i}(r)
		    -  U_{n_vlj}^{N_i}(r)\bigr] \nonumber\\
              &-&   \sum_{n_c,n_{c'}}
                   \widetilde{\bf P}_{n_clj}^{N_i}
                   \biggl[\frac{U_{n_clj}^{N_i}(r)+U_{n_{c'}lj}^{N_i}(r)}{2}
		      -   U_{n_vlj}^{N_i}(r)\biggr]
                   \widetilde{\bf P}_{n_{c'}lj}^{N_i}\Bigr\},
\end{eqnarray}
 where
\[
  {\bf P}_{lj} = \sum_{m=-j}^j
    \bigl| ljm \bigl\rangle \bigr\langle ljm \bigr|,
\]
\[
  \widetilde{\bf P}_{n_clj}^{N_i} = \sum_{m=-j}^j
  \bigl| (\widetilde{n_cljm})^{N_i} \bigl\rangle 
  \bigr\langle (\widetilde{n_cljm})^{N_i} \bigr|,
\]
 $\bigl| ljm \bigl\rangle \bigr\langle ljm \bigr|$
 is the projector on the two-component spin-angular function $\chi_{ljm}$,
 $\bigl| (\widetilde{n_cljm})^{N_i} \bigl\rangle 
 \bigr\langle (\widetilde{n_cljm})^{N_i} \bigr|$
 is the projector on the outer core pseudospinor
 $\widetilde{\varphi}_{n_clj}^{N_i}\chi_{ljm}$, 
 $U_{n_vlj}^{N_i}$ and $U_{n_clj}^{N_i}$ are the radial components 
 of the GRECP derived for valence $\widetilde{\varphi}_{n_vlj}^{N_i}$
 and outer core $\widetilde{\varphi}_{n_clj}^{N_i}$ pseudospinors
 for the OMC $d$ or $f$ shell occupation number $N_i\ (i{=}1,2)$,
 $E_{\rm core}^{N_i}$ is the core energy, 
 $L$ is one more than the highest orbital angular momentum of the inner core 
 spinors and $J=L+1/2$.
 The separable terms (the second and third lines in Eq.~(\ref{UGRECP}))
 are added to the conventional radially-local RECP operator. These terms take
 into account the difference between the
 potentials acting on the outer core and valence electrons with the same
 $l$ and $j$.
\item
     The self-consistent GRECP,
     ${\bf U}^{\rm SfC}$, with the quadratic correction writes as
\begin{eqnarray}
      {\bf U}^{\rm SfC}  &=&
                   \frac{{\bf U}^{N_1}{+}{\bf U}^{N_2}}{2}
                 + \frac{{\bf U}^{N_1}{-}{\bf U}^{N_2}}{N_1-N_2}
                   \biggl(N_{\rm omc} - \frac{N_1{+}N_2}{2}\biggr)\nonumber\\
                 &+& B \biggl(N_{\rm omc} - \frac{N_1{+}N_2}{2}\biggr)^2\ ,
 \label{SfC_RECP}
\end{eqnarray}
     where $B$ is some adjustable parameter, $N_{\rm omc}{=}
     \langle\tilde{\Psi}|{\bf N}_{\rm omc}|\tilde{\Psi}\rangle$, 
     $\tilde{\Psi}$ is the many-electron wavefunction for the calculated state,
     and ${\bf N}_{\rm omc}$ is the occupation number operator of the 
     considered $d$ ($f$) shell that is written as

\begin{equation}
   {\bf N}_{\rm omc} = \sum_{j=|l-1/2|}^{l+1/2} \sum_{m=-j}^j
             \tilde{\bf a}_{n_{\rm omc}l_{\rm omc}jm}^{\dagger}
             \tilde{\bf a}_{n_{\rm omc}l_{\rm omc}jm}\ ,
 \label{Occ_num}
\end{equation}
      $\tilde{\bf a}_{n_{\rm omc}l_{\rm omc}jm}^{\dagger}$
     ($\tilde{\bf a}_{n_{\rm omc}l_{\rm omc}jm}$) is the creation
     (annihilation) operator for the electron in the pseudostate
     $|\widetilde{n_{\rm omc}l_{\rm omc}jm}\rangle$ corresponding the
     original one-electron state $|n_{\rm omc}l_{\rm omc}jm\rangle$,
     $n_{\rm omc}$ and $l_{\rm omc}$ are the principal and orbital quantum
     numbers of the OMC shell.
\item
The $\widetilde{\bf P}_{n_clj}^{N_i}$ projectors in ${\bf U}^{N_i}$ 
from Eq.~(\ref{SfC_RECP}) are replaced by the projectors 
\[
  \widetilde{\bf P}_{n_clj}^{\rm av} = \sum_{m=-j}^j
  \bigl| (\widetilde{n_cljm})^{\rm av} \bigl\rangle 
  \bigr\langle (\widetilde{n_cljm})^{\rm av} \bigr|
\]
 for simplicity, where
 $\bigl| (\widetilde{n_cljm})^{\rm av} \bigl\rangle 
 \bigr\langle (\widetilde{n_cljm})^{\rm av} \bigr|$
 is the projector on the outer core 
 pseudospinor
 $\widetilde{\varphi}_{n_clj}^{\rm av}\chi_{ljm}$, 

\begin{equation}
 \widetilde{\varphi}_{n_clj}^{\rm av}(r) = C_{\rm norm}
  \bigl[\widetilde{\varphi}_{n_clj}^{\rm N_1}(r)
  + \widetilde{\varphi}_{n_clj}^{\rm N_2}(r)\bigr]\ ,
\end{equation}
 and $C_{\rm norm}$ is the normalizing factor.
\end{enumerate}
 The comparison of self-consistent and conventional GRECP versions by 
 accuracy in calculations on the uranium and plutonium atoms can be found 
 in paper \cite{Petrov:04b}.

\subsection{Term-splitting correction for $d,f$-elements}
 \label{sTScor}

 The self-consistent (G)RECP correction gives no improvement in description of
 splittings to terms, e.g., of the configuration $5f_{5/2}^3 6d_{3/2}^1
 7s_{1/2}^2$ of uranium as compared to the parent (G)RECPs
 \cite{Titov:99,Titov:00a}.  Analysis of the corresponding errors shows that the
 main contribution (about 90~\%) is due to smoothing the original OMC spinors
 in the core region.
 The simplest way to minimize these errors is to use such (G)RECPs, in which
 the $5f$ shell is described by nodal pseudospinors, whereas the $4f$
 pseudospinors are nodeless. To reduce computational efforts, the $4f$ shell
 can be treated as ``frozen'' using the level-shift technique
 \cite{Titov:99,Titov:01}.

 If the small magnitude of the OMC shell ($5f$ here) relaxation is taken into
 account, there is another way out that can be optimal for the low-lying
 states.  It was suggested in \cite{Titov:99} to add the Term-Splitting (TS)
 correction (see also \cite{Titov:00a}) to the (G)RECP operator
\begin{eqnarray}
  {\bf U}^{\rm TS} &=& \sum_{x_1,x_2,x_3,x_4}  \lambda_{x_1x_2,x_3x_4}
   \widetilde{\bigl| x_1 \bigl\rangle}
   \widetilde{\bigl| x_3 \bigl\rangle}
      \widetilde{\bigr\langle x_2 \bigr|}
      \widetilde{\bigr\langle x_4 \bigr|}
 \nonumber\\
 &-&2  \sum_w \sum_{x_1,x_2,x_3}  (\lambda_{x_1x_2,x_3x_3}-
 \lambda_{x_1x_3,x_3x_2}) \delta_{wx_3}
   \widetilde{\bigl| x_1 \bigl\rangle} \widetilde{\bigr\langle x_2 \bigr|}\ ,
 \label{SO-c1}
\end{eqnarray}
 where $\lambda_{x_1x_2,x_3x_4}$ is the difference between the two-electron
 integrals calculated with original spinors and pseudospinors for the generator
 state,
  the indices $w \equiv (n_{\rm occ}l_{\rm occ}j_{\rm occ}m_{\rm occ})$
  correspond to the occupied spinors for the calculated state, the indices
  $x \equiv (n_{\rm omc}l_{\rm omc}jm)$ 
 run over all possible $j=|l_{\rm omc}\pm1/2|$ and
 $m=-j,-j+1,\ldots j$ for the given OMC shell.  These terms correct the one-
 and two-electron integrals containing only the $5f$ pseudospinors
 of uranium in the considered case.

 \subsection{Accounting for the Breit interaction between different shells}
 \label{sDCBsh}

 Let us analyze contributions of the Breit interaction between electrons from
 different shells to the energy of a heavy atom \cite{Titov:02Dism}.
 We will use the estimate (e.g., see \cite{Labzowsky:93c})
\[
  \langle P,P' | ({\vec \alpha}_i{\cdot}{\vec \alpha}_{i'}) | P,P' \rangle \sim
      \frac{1}{c^2} \langle ({\vec v}_P{\cdot}{\vec v}_{P'}) \rangle\ ;
\]
 for an uncoupled one-electron state $P$:
$
  \langle P|\vec{\alpha}|P \rangle{\sim}\frac{\langle\vec{v}\rangle_P}{c} ,
  \frac{|\langle\vec{v}\rangle_P|}{c}{\sim}\alpha Z_P^* ,
$
 where $\vec{\alpha}_i$ are $4{\times}4$ Dirac 
 matrices for the $i$-th electron, $c$ and $\vec{v}$ are velocities of light
 and electron, $\alpha{\approx}\frac{1}{137}$ is the fine structure constant.
 In the above expression  a ``pseudocharge'', $Z_P^*$, is introduced which
 can be most naturally defined in our consideration
 as \cite{Petrov:04b}
\begin{equation}
   Z_P^* = \langle P|\frac{1}{r}|P \rangle\ ,
 \label{Zpse}
\end{equation}
 that coincides with the nuclear charges only for nonrelativistic electrons
 occupying the ground states in hydrogen-like ions.  Besides,
 $\langle\frac{1}{r_{12}}\rangle$ can be estimated as
 $\langle\frac{1}{r}\rangle$ for the outermost of the one-electron states
 $P,P'$ \cite{Titov:02Dism}:

\[
  \langle P,P'| \frac{1}{r_{12}} |P,P' \rangle \sim
     \min\left[ \langle P|\frac{1}{r}|P \rangle,
          \langle P'|\frac{1}{r}|P' \rangle \right] =
     \min\left[ Z_P^*, Z_{P'}^* \right]\ .
\]
 As a result, the Breit interaction between the one-electron states $P$ and
 $P'$ can be estimated as
\[
  B_{PP'} 
  \approx
  \alpha^2 Z_P^* Z_{P'}^* \cdot
         \min\left[ Z_P^*,Z_{P'}^* \right] \cdot {\cal F}
\]
 where the correcting factor ${\cal F} \sim [0.1 \div 1]$ is introduced, 
 which depends on $\Delta l{=}|l_P{-}l_{P'}|,\Delta j{=}|j_P{-}j_{P'}|$,
 etc.

 Applying Eq.~(\ref{Zpse}) for inner core ($P{\equiv}f$), outer core
 ($P{\equiv}c$) and valence ($P{\equiv}v$) electrons one has $Z_f^*{\sim}100$,
 $Z_c^*{\sim}3$, $Z_v^*{\sim}1$ by the order of magnitude ($Z_P^*$ differs from
 an ``effective charge'' of the core with respect to the electron in the $P$-th
 state, $Z_P^{\rm Ef}{=}Z{-}N_c^P$, that is usually used in RECP calculations,
 where $Z$ is the nuclear charge, $N_c^P$ is the number of core electrons with
 respect to the $P$-th state).
 Therefore, ${\cal B}_{PP'}{\equiv}{\cal F}^{-1} B_{PP'}$ is as
$$
\begin{array}{llrllrllr}
     {\cal B}_{ff'} & \sim & 10~000~000\ {\rm cm}^{-1}\ ,~~ &
     {\cal B}_{fc}  & \sim &   9~000\ {\rm cm}^{-1}\ ,~~ &
     {\cal B}_{fv}  & \sim & 1000\ {\rm cm}^{-1}\ , \\
      {\cal B}_{cf}  & \sim &   9~000\ {\rm cm}^{-1}\ ,~~ &
     {\cal B}_{cc'} & \sim &  270\ {\rm cm}^{-1}\ ,~~ &
     {\cal B}_{cv}  & \sim &   30\ {\rm cm}^{-1}\ ,\\
      {\cal B}_{vf}  & \sim & 1000\ {\rm cm}^{-1}\ ,~~ &
      {\cal B}_{vc}  & \sim &   30\ {\rm cm}^{-1}\ ,~~ &
     {\cal B}_{vv'} & \sim &   10\ {\rm cm}^{-1}\ .\\
\end{array}
$$

 Let us consider approximations in accounting for the Breit interaction, that
 we made when outer core and valence electrons are included in GRECP
 calculations with Coulomb two-electron interactions, but inner core electrons
 are absorbed into the GRECP.  When both electrons belong to the inner core
 shells, the Breit effect is of the same order as the Coulomb interaction
 between them.  Though $B_{ff'}$ does not contribute to ``differential''
 (valence) properties directly, it can lead to essential relaxation of both
 core and valence shells.  This relaxation is taken into account when the Breit
 interaction is treated by self-consistent way in the framework of the HFDB
 method \cite{Quiney:87, Lindroth:89c}.

 The inner core electrons occupy closed shells. The only exchange part of the
 two-electron Breit interaction between the valence, outer core and inner core
 electrons, $B_{fv}$ and $B_{fc}$, gives non-zero contribution.  The
 contributions from $B_{fv}$ and $B_{fc}$, are quite essential for calculation
 at the level of ``chemical accuracy'' (about 1~kcal/mol or 350~cm$^{-1}$ for
 transition energies).  This accuracy level is, in general, determined by the
 possibilities of modern correlation methods and computers already for
 compounds of light elements.
 Note, that the contribution from the exchange interaction is not smaller than
 that from the Coulomb part \cite{Petrov:04b}.  The inner core electrons can be
 considered as ``frozen'' in most physical-chemical processes of interest.
 Therefore, the effective operators for $B_{fv}$ and $B_{fc}$ acting on the
 valence and outercore shells, $B_{fv}^{\rm Ef}$ and $B_{fc}^{\rm Ef}$, are of
 the same kind as the exchange $f{-}v$ and $f{-}c$ contributions of the SCF
 field in the Huzinaga-type potential, i.e.\ these terms can be well
 approximated by the spin-dependent potential of the form:

$$ 
 B_{fv}^{\rm Ef}+B_{fc}^{\rm Ef} = \sum_{lj} V_{lj}^{Br}(r) {\bf P}_{lj}\ + 
                \sum_{n_clj}\bigl[V_{n_clj}^{Br}(r)
		-  V_{lj}^{Br}(r)\bigr]\widetilde{\bf P}_{n_clj}, 
$$
 which has basically the same spin-angular structure as the
 GRECP has.  Thus, it can be taken into account directly when the HFDB (not
 HFD) calculation \cite{Grant:00b} is performed to generate outer core and
 valence bispinors but in the inversion procedure of the HF equations for
 generating the components of GRECP, the conventional interelectronic Coulomb
 interaction should be used instead of the Coulomb-Breit one.  Then, in the
 GRECP calculations one should consider only the Coulomb interaction between
 the explicitly treated electrons. 

 Due to small relaxation of outer core shells in most
 processes of interest, these shells can be also considered as ``frozen'' when
 analyzing the Breit contributions and the $B_{cc'}$ and $B_{cv}$ terms can be
 taken into account similarly to the $B_{fc}$ and $B_{fv}$ ones.  The error of
 this approximation will be additionally suppressed by relative weakness of the
 Breit interaction with the outer core electrons as compared to the inner
 core ones.  We note here, that the estimates for $Z_c^*$, $Z_v^*$ and,
 therefore, for $B_{cc'}$, $B_{cv}$ and $B_{vv'}$ given above are rather the
 upper limits.  For heavy atoms these Breit contributions are smaller
 approximately by one--two orders of magnitude.
 This decrease is due to enlarged radii of the valence and outer core shells
 and other effects in heavy atoms \cite{Petrov:04b}.
 For example, for uranium ($Z=92$) one has
 $Z_{1s}^*[\rm nonrel.\,SCF]{\sim} 92.4$,
 $Z_{1s}^*[\rm DHFB]{\sim} 122.4$
 (starting from $Z \sim 30$, $Z_{1s}^*$ grows faster than $Z$ due to
 relativistic effects, whereas $Z_{nl}^*$ is essentially smaller than
 the corresponding effective charge $Z_{nl}^{\rm Ef}$ for all other $nl$),
 $Z_{5f}^*{\sim} 1$, $Z_{6s}^*{\sim} 1$, $Z_{6p}^*{\sim} 0.7$, 
 $Z_{6d}^*{\sim} 0.4$, $Z_{7s}^*{\sim} 0.3$.
 Thus, $B_{cc'}$, $B_{cv}$, and $B_{vv'}$ contributions are
 negligible for the ``chemical accuracy'' of calculation. 
 Therefore, the above made estimates provide us
 a good background for approximating
 the Breit interaction by a one-electron GRECP operator that should work well
 both for actinides and for superheavy elements.  The numerical tests of the
 GRECPs accounting for the Breit effects are discussed in the next section.

\section{Results and discussion}

 For all-electron calculations, we used the atomic HFDB code
 \cite{Bratzev:77,Tupitsyn:02A} which allows one to account for the Breit
 interactions both in the framework of the first-order perturbation theory
 (PT-1) and by the self-consistent way as well as to account for different
 models of nuclear charge distribution.  For test calculations with (G)RECPs,
 the atomic Hartree-Fock code in the $jj$-coupling scheme ({\sc hfj})
 \cite{Tupitsyn:95} was used (that was quite sufficient for studying errors of
 the one-electron (G)RECP operators).  Both the codes are numerical that allows
 us to exclude the errors due to the incompleteness of basis sets when
 estimating accuracy of different RECPs and GRECPs.

 The transition energies between states averaged over the low-lying
 configurations of SHEs 112, 113, 114 and actinides U, Pu are presented in
 tables \ref{E112_conf}, \ref{E113_conf}, \ref{E114_conf} and \ref{U_conf},
 \ref{Pu_conf}, respectively.  One can see that the errors due to the point
 nuclear model reach a few thousand wave numbers for the SHEs and several
 hundred wave numbers for the actinides.  The considered small variations in
 the nuclear charge distribution (including the nuclear size) in the framework
 of finite-size nuclei lead to change of the transition energies for the
 studied SHEs less than on 60~cm$^{-1}$.  The differences between the 
 results with the \mbox{PT-1}
 and self-consistent ways of accounting for the Breit interaction are
 within 7~cm$^{-1}$ for SHEs and actinides whereas neglecting the Breit effects
 leads to the errors up to a few thousand wave numbers for the studied
 actinides and several hundred wave numbers for the SHEs.

 The GRECP errors in reproducing the results of the all-electron HFDB
 calculations with the Fermi nuclear model are collected into two groups.
 First, the GRECP errors for transitions without change in the occupation
 number of the $6d$ shell for the SHEs (tables \ref{E112_conf}, \ref{E113_conf}
 and \ref{E114_conf}) and the $5f$ shell for the actinides (tables \ref{U_conf}
 and \ref{Pu_conf}) are relatively small whereas the corresponding errors of
 the other tested RECPs for the SHEs are significantly higher. The same number
 of electrons is explicitly treated in calculations with different (G)RECP
 versions for a given atom.  Here and further, we do not discuss the particular
 case of the 52e-GRECP for E112 if the opposite is not explicitly stated.  The
 RECPs of other groups for uranium were tested in paper \cite{Titov:99}. It
 should be noted that they do not take into account the large contribution from
 the Breit interaction.  The Breit effects were also not considered at the
 generation stage of the RECP of Nash {\it et al.} \cite{Nash:97}. However, it
 can not explain the large errors for this RECP in tables \ref{E112_conf},
 \ref{E113_conf} and \ref{E114_conf}. It is not clear from paper \cite{Nash:97}
 which nuclear model was used there. The Breit interaction was taken into
 account only in the \mbox{PT-1}
 approximation at the generation stage of the PseudoPotential (PP) of Seth
 {\it et al}.  However, the corresponding changes in the transition energies
 are negligible in comparison with the PP errors.

 Second, the GRECP errors for transitions with excitation of one $6d$ electron
 for the SHEs or one $5f$ electron for the actinides are about 400~cm$^{-1}$.
 These errors have a systematic nature (unlike the corresponding errors for the
 tested RECPs of other groups) and are connected with the fact that the OMC
 $6d$ shell for the SHEs and the OMC $5f$ shell for the actinides in the
 present GRECP versions are described with the help of nodeless pseudospinors.
 Obviously, these errors can be reduced significantly if one includes the
 $5d,5f$ electrons for the SHEs and the $4f$ electrons for the actinides
 explicitly in the GRECP calculations (see the 52e-GRECP results for E112 in
 table~\ref{E112_conf}).  The corresponding pseudospinors can be then
 ``frozen'' in these GRECP calculations with the help of the level-shift
 technique \cite{Titov:99,Titov:01} to reduce the computational efforts.
 Alternatively, the 
 self-consistent GRECP method described in section \ref{sUSfC} can be used.
 
 The energies of splittings between terms are considered in table
 \ref{E112_term} for E112 and table \ref{U_term} for U.  The errors of the RECP
 and GRECP approximations and the errors caused by neglecting the Breit effects
 are within 200~cm$^{-1}$ for E112 (except for the RECP of Nash {\it et al.}).
 The Breit contributions to the term-splitting energies for U are within
 100~cm$^{-1}$ whereas the GRECP errors are up to 750~cm$^{-1}$.  The latter 
 can be reduced drastically by applying the 
 term-splitting correction (see section \ref{sTScor} and table \ref{U_term}).
 The results show that addition of the 
 term-splitting correction allows one to reduce the most serious errors up to
 10 times for the splittings into terms, thus reducing the errors for the
 energies of transition between terms to the same order of magnitude as the
 errors for transitions between the states averaged over the configurations
 (when only the self-consistent GRECP is applied).
 Obviously, any transition between two different terms having different
 occupation numbers of the OMC shell, ${\bf N}_{\rm omc}^1$ and ${\bf N}_{\rm
 omc}^2$, can be presented as a combination of three consequent transitions:
 transiton from the first term to the average over the configuration with the
 same ${\bf N}_{\rm omc}^1$, transition between averages over configurations
 with ${\bf N}_{\rm omc}^1$ and ${\bf N}_{\rm omc}^2$ and transition from the
 latter to the second term with ${\bf N}_{\rm omc}^2$.
 Therefore, applying of both the self-consistent and term-spitting GRECP
 corrections to treatment of transitions between any terms
 allows one to reduce dramatically the (G)RECP approximation errors without
 increasing the number of explicitly treated core electrons of a considered
 $d,f$-element.

 In tables \ref{E112_r2} and \ref{E112_int}, the matrix elements of $<r^2>$ and
 radial integrals $2\int^{\infty}_{R_{n}} \mid f_{n_vlj}(r) [f_{n_vlj}(r) -
 \widetilde{\varphi}_{n_vlj}(r)] \mid dr$ (where $f_{n_vlj}$ is the large
 component of the Dirac spinor, $\widetilde{\varphi}_{n_vlj}$ is the radial
 part of the corresponding pseudospinor and $R_{n}$ is the radius of the last
 spinor node) are considered for the cases of spinors from different
 configurations of E112.  The errors in these matrix elements and integrals
 characterize the quality of reproducing the electronic density in outer core
 and valence regions of the atom. One can see that the GRECP allows one to
 reproduce the electronic density in the valence region (the $7s_{1/2}$ and
 $7p_{1/2}$ spinors) with very high accuracy.  The one-electron energies for
 spinors from different configurations of E112 are presented in table
 \ref{E112_orb}.  Similar conclusion can be made in the latter case.

\section*{Conclusions}

 Different nuclear models and contributions of the Breit interaction between
 valence, inner and outer core shells of uranium, plutonium and superheavy
 elements E112, E113, and E114 are considered in the framework of all-electron
 four-component and (G)RECP methods.  It is concluded on the basis of the
 performed calculations and theoretical analysis that the Breit contributions
 with inner core shells must be taken into account in calculations of
 actinide and SHE compounds with ``chemical accuracy'' whereas those between
 valence and outer core shells can be omitted.

 The differences in the atomic energies between the cases of the PT-1 and
 self-consistent ways of treating the Breit interaction as well as small
 variations in the nuclear charge distribution in the framework of finite-size
 nuclei are not essential for the considered accuracy of calculations.
 However, the difference between the point and finite nuclear models is
 important for the valence (transition) energies.  The effects of accounting
 for the Breit interaction and finite nuclear model can be simulated by GRECPs
 with very good accuracy when only Coulomb interaction between the explicitly
 treated electrons is taken into account.  Thus, the GRECP method allows one to
 carry out reliable calculations of actinides, SHEs and their compounds at the
 level of ``chemical accuracy''.

\section*{Acknowledgments}

 The present work is supported by the U.S.\ CRDF Grant No.\ RP2--2339--GA--02
 and the RFBR grant 03--03--32335.  N.M.\ is also supported by the grants of
 Russian Science Support Foundation and the governor of Leningrad district.
 A.P.\ is grateful to Ministry of education of Russian Federation (grant
 PD\,02--1.3--236).

\bibliographystyle{apsrev}            

\bibliography{bib/JournAbbr,bib/Titov,bib/TitovLib,bib/Kaldor,bib/Isaev,bib/AbsConf/TitovAbs}

\newcommand{\noopsort}[1]{} \newcommand{\printfirst}[2]{#1}
  \newcommand{\singleletter}[1]{#1} \newcommand{\switchargs}[2]{#2#1}
\begin{thebibliography}{37}
\expandafter\ifx\csname natexlab\endcsname\relax\def\natexlab#1{#1}\fi
\expandafter\ifx\csname bibnamefont\endcsname\relax
  \def\bibnamefont#1{#1}\fi
\expandafter\ifx\csname bibfnamefont\endcsname\relax
  \def\bibfnamefont#1{#1}\fi
\expandafter\ifx\csname citenamefont\endcsname\relax
  \def\citenamefont#1{#1}\fi
\expandafter\ifx\csname url\endcsname\relax
  \def\url#1{\texttt{#1}}\fi
\expandafter\ifx\csname urlprefix\endcsname\relax\def\urlprefix{URL }\fi
\providecommand{\bibinfo}[2]{#2}
\providecommand{\eprint}[2][]{\url{#2}}

\bibitem[{\citenamefont{Hoffman and {M\"nzenberg}}(2000)}]{Hoffman:00}
\bibinfo{author}{\bibfnamefont{S.}~\bibnamefont{Hoffman}} \bibnamefont{and}
  \bibinfo{author}{\bibfnamefont{G.}~\bibnamefont{{M\"nzenberg}}},
  \bibinfo{journal}{Rev.\ Mod.\ Phys.} \textbf{\bibinfo{volume}{72}},
  \bibinfo{pages}{733} (\bibinfo{year}{2000}).

\bibitem[{\citenamefont{Oganessian et~al.}(1999)}]{Oganessian:99}
\bibinfo{author}{\bibfnamefont{Y.~T.} \bibnamefont{Oganessian}}
  \bibnamefont{et~al.}, \bibinfo{journal}{Nature}
  \textbf{\bibinfo{volume}{400}}, \bibinfo{pages}{242} (\bibinfo{year}{1999}).

\bibitem[{\citenamefont{Oganessian}(2001)}]{Oganessian:01}
\bibinfo{author}{\bibfnamefont{Y.}~\bibnamefont{Oganessian}},
  \bibinfo{journal}{Nature} \textbf{\bibinfo{volume}{413}},
  \bibinfo{pages}{122} (\bibinfo{year}{2001}).

\bibitem[{\citenamefont{{Sch\"adel}}(2003)}]{Schadel:03b}
\bibinfo{editor}{\bibfnamefont{M.}~\bibnamefont{{Sch\"adel}}}, ed.,
  \emph{\bibinfo{title}{The Chemistry of Superheavy Elements}}
  (\bibinfo{publisher}{Kluwer}, \bibinfo{address}{Dordrecht},
  \bibinfo{year}{2003}), \bibinfo{note}{318~pp.}

\bibitem[{\citenamefont{Hirao and Ishikawa}(2004)}]{Hirao:04}
\bibinfo{editor}{\bibfnamefont{K.}~\bibnamefont{Hirao}} \bibnamefont{and}
  \bibinfo{editor}{\bibfnamefont{Y.}~\bibnamefont{Ishikawa}}, eds.,
  \emph{\bibinfo{title}{Recent Advances in Relativistic Molecular Theory}}
  (\bibinfo{publisher}{World Scientific}, \bibinfo{address}{Singapore},
  \bibinfo{year}{2004}), \bibinfo{note}{{328\,pp}}.

\bibitem[{\citenamefont{Schwerdtfeger}(2004)}]{Schwerdtfeger:04aa}
\bibinfo{editor}{\bibfnamefont{P.}~\bibnamefont{Schwerdtfeger}}, ed.,
  \emph{\bibinfo{title}{Relativistic Electronic Structure Theory. {P}art~2.
  {A}pplications}}, vol.~\bibinfo{volume}{14} of
  \emph{\bibinfo{series}{Theoretical and Computational Chemistry}}
  (\bibinfo{publisher}{Elsevier}, \bibinfo{address}{Amsterdam},
  \bibinfo{year}{2004}), \bibinfo{note}{{xv\,+\,787\,pp}}.

\bibitem[{\citenamefont{Mohr}(1997)}]{Mohr:97}
\bibinfo{author}{\bibfnamefont{P.~J.} \bibnamefont{Mohr}},
  \bibinfo{journal}{Phys.\ Rep.} \textbf{\bibinfo{volume}{293}},
  \bibinfo{pages}{227} (\bibinfo{year}{1997}).

\bibitem[{\citenamefont{Grant and Quiney}(2000)}]{Grant:00b}
\bibinfo{author}{\bibfnamefont{I.~P.} \bibnamefont{Grant}} \bibnamefont{and}
  \bibinfo{author}{\bibfnamefont{H.~M.} \bibnamefont{Quiney}},
  \bibinfo{journal}{Int.\ J.\ Quantum Chem.} \textbf{\bibinfo{volume}{80}},
  \bibinfo{pages}{283} (\bibinfo{year}{2000}).

\bibitem[{\citenamefont{Reiher and Hess}(2000)}]{Reiher:00}
\bibinfo{author}{\bibfnamefont{M.}~\bibnamefont{Reiher}} \bibnamefont{and}
  \bibinfo{author}{\bibfnamefont{B.~A.} \bibnamefont{Hess}}, in
  \emph{\bibinfo{booktitle}{Modern Methods and Algorithms of Quantum
  Chemistry}}, edited by
  \bibinfo{editor}{\bibfnamefont{J.}~\bibnamefont{Grotendorst}}
  (\bibinfo{address}{J{\"u}lich}, \bibinfo{year}{2000}),
  vol.~\bibinfo{volume}{1}, pp. \bibinfo{pages}{451--477},
  \bibinfo{note}{{[http://www.fz-juelich.de/nic-series]}}.

\bibitem[{\citenamefont{Shabaev}(2002)}]{Shabaev:02a}
\bibinfo{author}{\bibfnamefont{V.~M.} \bibnamefont{Shabaev}},
  \bibinfo{journal}{Phys.\ Rep.} \textbf{\bibinfo{volume}{356}},
  \bibinfo{pages}{119} (\bibinfo{year}{2002}).

\bibitem[{\citenamefont{Labzowsky and Goidenko}(2002)}]{Labzowsky:02b}
\bibinfo{author}{\bibfnamefont{L.~N.} \bibnamefont{Labzowsky}}
  \bibnamefont{and} \bibinfo{author}{\bibfnamefont{I.}~\bibnamefont{Goidenko}},
  in \emph{\bibinfo{booktitle}{Relativistic Electronic Structure Theory.
  {Part~I}. {F}undamentals}}, edited by
  \bibinfo{editor}{\bibfnamefont{P.}~\bibnamefont{Schwerdtfeger}}
  (\bibinfo{organization}{Elsevier}, \bibinfo{address}{Amsterdam},
  \bibinfo{year}{2002}), pp. \bibinfo{pages}{401--467}.

\bibitem[{\citenamefont{Visscher}(1996)}]{Visscher:96}
\bibinfo{author}{\bibfnamefont{L.}~\bibnamefont{Visscher}},
  \bibinfo{journal}{Chem.\ Phys.\ Lett.} \textbf{\bibinfo{volume}{253}},
  \bibinfo{pages}{20} (\bibinfo{year}{1996}).

\bibitem[{\citenamefont{Dyall}(2002)}]{Dyall:02a}
\bibinfo{author}{\bibfnamefont{K.~G.} \bibnamefont{Dyall}},
  \bibinfo{journal}{J.\ Comput.\ Chem.} \textbf{\bibinfo{volume}{23}},
  \bibinfo{pages}{786} (\bibinfo{year}{2002}).

\bibitem[{\citenamefont{Visscher}(2002)}]{Visscher:02aa}
\bibinfo{author}{\bibfnamefont{L.}~\bibnamefont{Visscher}},
  \bibinfo{journal}{J.\ Comput.\ Chem.} \textbf{\bibinfo{volume}{23}},
  \bibinfo{pages}{759} (\bibinfo{year}{2002}).

\bibitem[{\citenamefont{Ermler et~al.}(1988)\citenamefont{Ermler, Ross, and
  Christiansen}}]{Ermler:88}
\bibinfo{author}{\bibfnamefont{W.~C.} \bibnamefont{Ermler}},
  \bibinfo{author}{\bibfnamefont{R.~B.} \bibnamefont{Ross}}, \bibnamefont{and}
  \bibinfo{author}{\bibfnamefont{P.~A.} \bibnamefont{Christiansen}},
  \bibinfo{journal}{Adv.\ Quantum Chem.} \textbf{\bibinfo{volume}{19}},
  \bibinfo{pages}{139} (\bibinfo{year}{1988}).

\bibitem[{\citenamefont{Titov et~al.}(2005)\citenamefont{Titov, Mosyagin,
  Petrov, and Isaev}}]{Titov:05b}
\bibinfo{author}{\bibfnamefont{A.~V.} \bibnamefont{Titov}},
  \bibinfo{author}{\bibfnamefont{N.~S.} \bibnamefont{Mosyagin}},
  \bibinfo{author}{\bibfnamefont{A.~N.} \bibnamefont{Petrov}},
  \bibnamefont{and} \bibinfo{author}{\bibfnamefont{T.~A.} \bibnamefont{Isaev}},
  \bibinfo{journal}{Progr.\ Theor.\ Chem.\ Phys.}  (\bibinfo{year}{2005}),
  \bibinfo{note}{in press}.

\bibitem[{\citenamefont{Tupitsyn et~al.}(1995)\citenamefont{Tupitsyn, Mosyagin,
  and Titov}}]{Tupitsyn:95}
\bibinfo{author}{\bibfnamefont{I.~I.} \bibnamefont{Tupitsyn}},
  \bibinfo{author}{\bibfnamefont{N.~S.} \bibnamefont{Mosyagin}},
  \bibnamefont{and} \bibinfo{author}{\bibfnamefont{A.~V.} \bibnamefont{Titov}},
  \bibinfo{journal}{J.\ Chem.\ Phys.} \textbf{\bibinfo{volume}{103}},
  \bibinfo{pages}{6548} (\bibinfo{year}{1995}).

\bibitem[{\citenamefont{Mosyagin et~al.}(1997)\citenamefont{Mosyagin, Titov,
  and Latajka}}]{Mosyagin:97}
\bibinfo{author}{\bibfnamefont{N.~S.} \bibnamefont{Mosyagin}},
  \bibinfo{author}{\bibfnamefont{A.~V.} \bibnamefont{Titov}}, \bibnamefont{and}
  \bibinfo{author}{\bibfnamefont{Z.}~\bibnamefont{Latajka}},
  \bibinfo{journal}{Int.\ J.\ Quantum Chem.} \textbf{\bibinfo{volume}{63}},
  \bibinfo{pages}{1107} (\bibinfo{year}{1997}).

\bibitem[{\citenamefont{Titov and Mosyagin}(1999)}]{Titov:99}
\bibinfo{author}{\bibfnamefont{A.~V.} \bibnamefont{Titov}} \bibnamefont{and}
  \bibinfo{author}{\bibfnamefont{N.~S.} \bibnamefont{Mosyagin}},
  \bibinfo{journal}{Int.\ J.\ Quantum Chem.} \textbf{\bibinfo{volume}{71}},
  \bibinfo{pages}{359} (\bibinfo{year}{1999}).

\bibitem[{\citenamefont{Phillips and Kleinman}(1959)}]{Phillips:59}
\bibinfo{author}{\bibfnamefont{J.~C.} \bibnamefont{Phillips}} \bibnamefont{and}
  \bibinfo{author}{\bibfnamefont{L.}~\bibnamefont{Kleinman}},
  \bibinfo{journal}{Phys.\ Rev.} \textbf{\bibinfo{volume}{116}},
  \bibinfo{pages}{287} (\bibinfo{year}{1959}).

\bibitem[{\citenamefont{Abarenkov and Heine}(1965)}]{Abarenkov:65}
\bibinfo{author}{\bibfnamefont{I.~V.} \bibnamefont{Abarenkov}}
  \bibnamefont{and} \bibinfo{author}{\bibfnamefont{V.}~\bibnamefont{Heine}},
  \bibinfo{journal}{Philos.\ Mag.} \textbf{\bibinfo{volume}{12}},
  \bibinfo{pages}{529} (\bibinfo{year}{1965}).

\bibitem[{\citenamefont{Heine and Abarenkov}(1964)}]{Heine:64}
\bibinfo{author}{\bibfnamefont{V.}~\bibnamefont{Heine}} \bibnamefont{and}
  \bibinfo{author}{\bibfnamefont{I.~V.} \bibnamefont{Abarenkov}},
  \bibinfo{journal}{Philos.\ Mag.} \textbf{\bibinfo{volume}{9}},
  \bibinfo{pages}{451} (\bibinfo{year}{1964}).

\bibitem[{\citenamefont{Titov and Mosyagin}(1995)}]{Titov:95}
\bibinfo{author}{\bibfnamefont{A.~V.} \bibnamefont{Titov}} \bibnamefont{and}
  \bibinfo{author}{\bibfnamefont{N.~S.} \bibnamefont{Mosyagin}},
  \bibinfo{journal}{Structural Chem.} \textbf{\bibinfo{volume}{6}},
  \bibinfo{pages}{317} (\bibinfo{year}{1995}).

\bibitem[{\citenamefont{Titov and Mosyagin}(2000)}]{Titov:00a}
\bibinfo{author}{\bibfnamefont{A.~V.} \bibnamefont{Titov}} \bibnamefont{and}
  \bibinfo{author}{\bibfnamefont{N.~S.} \bibnamefont{Mosyagin}},
  \bibinfo{journal}{Russ.\ J.\ Phys.\ Chem.} \textbf{\bibinfo{volume}{74, {\rm
  Suppl.\,2}}}, \bibinfo{pages}{S376} (\bibinfo{year}{2000}),
  \bibinfo{note}{[arXiv: physics/0008160]}.

\bibitem[{\citenamefont{Mosyagin et~al.}(2000)\citenamefont{Mosyagin, Eliav,
  Titov, and Kaldor}}]{Mosyagin:00}
\bibinfo{author}{\bibfnamefont{N.~S.} \bibnamefont{Mosyagin}},
  \bibinfo{author}{\bibfnamefont{E.}~\bibnamefont{Eliav}},
  \bibinfo{author}{\bibfnamefont{A.~V.} \bibnamefont{Titov}}, \bibnamefont{and}
  \bibinfo{author}{\bibfnamefont{U.}~\bibnamefont{Kaldor}},
  \bibinfo{journal}{J.\ Phys.\ B} \textbf{\bibinfo{volume}{33}},
  \bibinfo{pages}{667} (\bibinfo{year}{2000}).

\bibitem[{\citenamefont{Isaev et~al.}(2000)\citenamefont{Isaev, Mosyagin,
  Kozlov, Titov, Eliav, and Kaldor}}]{Isaev:00}
\bibinfo{author}{\bibfnamefont{T.~A.} \bibnamefont{Isaev}},
  \bibinfo{author}{\bibfnamefont{N.~S.} \bibnamefont{Mosyagin}},
  \bibinfo{author}{\bibfnamefont{M.~G.} \bibnamefont{Kozlov}},
  \bibinfo{author}{\bibfnamefont{A.~V.} \bibnamefont{Titov}},
  \bibinfo{author}{\bibfnamefont{E.}~\bibnamefont{Eliav}}, \bibnamefont{and}
  \bibinfo{author}{\bibfnamefont{U.}~\bibnamefont{Kaldor}},
  \bibinfo{journal}{J.\ Phys.\ B} \textbf{\bibinfo{volume}{33}},
  \bibinfo{pages}{5139} (\bibinfo{year}{2000}).

\bibitem[{\citenamefont{Titov}()}]{Titov:02Dism}
\bibinfo{author}{\bibfnamefont{A.~V.} \bibnamefont{Titov}}, \bibinfo{note}{{\it
  Doctorate Thesis}, {(Petersburg Nuclear Physics Institute, RAS, Russia,
  2002)}}.

\bibitem[{\citenamefont{Titov et~al.}(1991)\citenamefont{Titov, Mitrushenkov,
  and Tupitsyn}}]{Titov:91}
\bibinfo{author}{\bibfnamefont{A.~V.} \bibnamefont{Titov}},
  \bibinfo{author}{\bibfnamefont{A.~O.} \bibnamefont{Mitrushenkov}},
  \bibnamefont{and} \bibinfo{author}{\bibfnamefont{I.~I.}
  \bibnamefont{Tupitsyn}}, \bibinfo{journal}{Chem.\ Phys.\ Lett.}
  \textbf{\bibinfo{volume}{185}}, \bibinfo{pages}{330} (\bibinfo{year}{1991}).

\bibitem[{\citenamefont{Petrov et~al.}(2004)\citenamefont{Petrov, Mosyagin,
  Titov, and Tupitsyn}}]{Petrov:04b}
\bibinfo{author}{\bibfnamefont{A.~N.} \bibnamefont{Petrov}},
  \bibinfo{author}{\bibfnamefont{N.~S.} \bibnamefont{Mosyagin}},
  \bibinfo{author}{\bibfnamefont{A.~V.} \bibnamefont{Titov}}, \bibnamefont{and}
  \bibinfo{author}{\bibfnamefont{I.~I.} \bibnamefont{Tupitsyn}},
  \bibinfo{journal}{J.\ Phys.\ B} \textbf{\bibinfo{volume}{37}},
  \bibinfo{pages}{4621} (\bibinfo{year}{2004}).

\bibitem[{\citenamefont{Mosyagin and Titov}()}]{Mosyagin:04a}
\bibinfo{author}{\bibfnamefont{N.~S.} \bibnamefont{Mosyagin}} \bibnamefont{and}
  \bibinfo{author}{\bibfnamefont{A.~V.} \bibnamefont{Titov}},
  \bibinfo{note}{arXiv.org/ physics/0406143 (2004); J.\ Chem.\ Phys., in press
  (2005)}.

\bibitem[{\citenamefont{Titov et~al.}(2001)\citenamefont{Titov, Mosyagin,
  Alekseyev, and Buenker}}]{Titov:01}
\bibinfo{author}{\bibfnamefont{A.~V.} \bibnamefont{Titov}},
  \bibinfo{author}{\bibfnamefont{N.~S.} \bibnamefont{Mosyagin}},
  \bibinfo{author}{\bibfnamefont{A.~B.} \bibnamefont{Alekseyev}},
  \bibnamefont{and} \bibinfo{author}{\bibfnamefont{R.~J.}
  \bibnamefont{Buenker}}, \bibinfo{journal}{Int.\ J.\ Quantum Chem.}
  \textbf{\bibinfo{volume}{81}}, \bibinfo{pages}{409} (\bibinfo{year}{2001}).

\bibitem[{\citenamefont{Labzowsky et~al.}(1993)\citenamefont{Labzowsky,
  Klimchitskaya, and Dmitriev}}]{Labzowsky:93c}
\bibinfo{author}{\bibfnamefont{L.~N.} \bibnamefont{Labzowsky}},
  \bibinfo{author}{\bibfnamefont{G.~L.} \bibnamefont{Klimchitskaya}},
  \bibnamefont{and} \bibinfo{author}{\bibfnamefont{Y.~Y.}
  \bibnamefont{Dmitriev}}, \emph{\bibinfo{title}{Relativistic Effects in the
  Spectra of Atomic Systems}} (\bibinfo{publisher}{Institute of Physics
  Publishing}, \bibinfo{address}{Bristol and Philadelphia},
  \bibinfo{year}{1993}), \bibinfo{note}{340 pp}.

\bibitem[{\citenamefont{Quiney et~al.}(1987)\citenamefont{Quiney, Grant, and
  Wilson}}]{Quiney:87}
\bibinfo{author}{\bibfnamefont{H.~M.} \bibnamefont{Quiney}},
  \bibinfo{author}{\bibfnamefont{I.~P.} \bibnamefont{Grant}}, \bibnamefont{and}
  \bibinfo{author}{\bibfnamefont{S.}~\bibnamefont{Wilson}},
  \bibinfo{journal}{J.\ Phys.\ B} \textbf{\bibinfo{volume}{20}},
  \bibinfo{pages}{1413} (\bibinfo{year}{1987}).

\bibitem[{\citenamefont{Lindroth et~al.}(1989)\citenamefont{Lindroth,
  {M{\aa}rtensson-Pendrill}, Ynnerman, and {\"Oster}}}]{Lindroth:89c}
\bibinfo{author}{\bibfnamefont{E.}~\bibnamefont{Lindroth}},
  \bibinfo{author}{\bibfnamefont{A.-M.}
  \bibnamefont{{M{\aa}rtensson-Pendrill}}},
  \bibinfo{author}{\bibfnamefont{A.}~\bibnamefont{Ynnerman}}, \bibnamefont{and}
  \bibinfo{author}{\bibfnamefont{P.}~\bibnamefont{{\"Oster}}},
  \bibinfo{journal}{J.\ Phys.\ B} \textbf{\bibinfo{volume}{22}},
  \bibinfo{pages}{2447} (\bibinfo{year}{1989}).

\bibitem[{\citenamefont{Bratzev et~al.}(1977)\citenamefont{Bratzev, Deyneka,
  and Tupitsyn}}]{Bratzev:77}
\bibinfo{author}{\bibfnamefont{V.~F.} \bibnamefont{Bratzev}},
  \bibinfo{author}{\bibfnamefont{G.~B.} \bibnamefont{Deyneka}},
  \bibnamefont{and} \bibinfo{author}{\bibfnamefont{I.~I.}
  \bibnamefont{Tupitsyn}}, \bibinfo{journal}{Bull.\ Acad.\ Sci.\ USSR, Phys.\
  Ser.} \textbf{\bibinfo{volume}{41}}, \bibinfo{pages}{173}
  (\bibinfo{year}{1977}).

\bibitem[{\citenamefont{Tupitsyn and Petrov}(2002)}]{Tupitsyn:02A}
\bibinfo{author}{\bibfnamefont{I.~I.} \bibnamefont{Tupitsyn}} \bibnamefont{and}
  \bibinfo{author}{\bibfnamefont{A.~N.} \bibnamefont{Petrov}}, in
  \emph{\bibinfo{booktitle}{5--th Session of the V.A. Fock School on Quantum
  and Computational Chemistry}} (\bibinfo{address}{Novgorod the Great},
  \bibinfo{year}{2002}), p.~\bibinfo{pages}{62}.

\bibitem[{\citenamefont{Nash et~al.}(1997)\citenamefont{Nash, Bursten, and
  Ermler}}]{Nash:97}
\bibinfo{author}{\bibfnamefont{C.~S.} \bibnamefont{Nash}},
  \bibinfo{author}{\bibfnamefont{B.~E.} \bibnamefont{Bursten}},
  \bibnamefont{and} \bibinfo{author}{\bibfnamefont{W.~C.}
  \bibnamefont{Ermler}}, \bibinfo{journal}{J.\ Chem.\ Phys.}
  \textbf{\bibinfo{volume}{106}}, \bibinfo{pages}{5133} (\bibinfo{year}{1997}),
  \bibinfo{note}{{[Erratum: JCP 111 (1999) 2347]}}.

\end{thebibliography}

\begin{table}
\caption{
  Transition energies (TE) between states averaged over the relativistic 
  configurations of E112 derived from
  HFDB calculations with Fermi nuclear model
  and the corresponding absolute errors of all-electron and (G)RECP
  calculations (in~cm$^{-1}$).
}
\label{E112_conf}
\begin{center}
\begin{tabular}{lrrrrrrrrrrr}
\hline
\hline
                                                          &  HFDB        & HFDB          &    HFDB           &   HFDB          &  HFD+B     & HFD       &      52e-   &      20e-   & Ionic     & 20e-RECP     & 20e-PP       \\
                                                          &  (Fermi,     & (Ball,        &    (Fermi,        &   (Point)       &  (Fermi,   & (Fermi,   &  GRECP      &  GRECP      & 20e-      & of Nash      & of Seth      \\
                                                          &   A=296)     &  A=296)       &    A=285)         &                 &   A=296)   &  A=296)   &             &             & RECP      & {\it et al.} & {\it et al.} \\
                                                          & (a)          & (b)           &    (a)            &   (c)           & (d)        & (e)       & (f)         & (f)         & (g)       & (h)          & (i)          \\
\hline                                                                                                                                                                                                                
 Configuration                                            & TE           &\multicolumn{10}{c}{  Absolute errors}                                                                                                              \\
\hline                                                                                                                                                                                                           
$6d_{3/2}^4 6d_{5/2}^6 7s_{1/2}^2 \rightarrow           $ &              &               &                   &                 &            &            &            &             &           &              &              \\   
$6d_{3/2}^4 6d_{5/2}^6 7s_{1/2}^1 7p_{1/2}^1            $ &       46406  &         -3    &            22     &         1768    &          1 &       -27  &        1   &        -17  &       588 &      3198    &     153      \\
$6d_{3/2}^4 6d_{5/2}^6 7s_{1/2}^1 7p_{3/2}^1            $ &       64559  &         -4    &            25     &         1964    &         -1 &       239  &        4   &        -29  &       820 &      5480    &      27      \\
$6d_{3/2}^4 6d_{5/2}^6 7s_{1/2}^1 8s_{1/2}^1            $ &       72571  &         -3    &            22     &         1760    &         -1 &       257  &        6   &        -25  &       719 &      5085    &     105      \\
$6d_{3/2}^4 6d_{5/2}^6 7s_{1/2}^1 7d_{3/2}^1            $ &       81845  &         -4    &            23     &         1879    &         -1 &       277  &        6   &        -18  &       809 &      5465    &      99      \\
\hline
\hline                                                                                                                                                                                                                                  
$6d_{3/2}^4 6d_{5/2}^5 7s_{1/2}^2 7p_{1/2}^1            $ &       28701  &          1    &            -8     &         -644    &          2 &      -576  &       31   &        305  &      -422 &     -3723    &     380      \\
$6d_{3/2}^4 6d_{5/2}^5 7s_{1/2}^2 7p_{3/2}^1            $ &       52595  &          1    &            -6     &         -464    &          0 &      -267  &       37   &        277  &      -181 &     -1254    &     189      \\
$6d_{3/2}^4 6d_{5/2}^5 7s_{1/2}^2 8s_{1/2}^1            $ &       62635  &          2    &           -10     &         -776    &          0 &      -252  &       43   &        314  &      -315 &     -1879    &     326      \\
$6d_{3/2}^4 6d_{5/2}^5 7s_{1/2}^2 7d_{3/2}^1            $ &       72443  &          1    &            -9     &         -666    &          0 &      -234  &       43   &        322  &      -220 &     -1514    &     314      \\
$6d_{3/2}^4 6d_{5/2}^5 7s_{1/2}^2                       $ &       84449  &          1    &            -9     &         -672    &          0 &      -234  &       43   &        322  &      -224 &     -1531    &     308      \\
\hline                                                                                                                                                                                                                                  
$6d_{3/2}^3 6d_{5/2}^6 7s_{1/2}^2 7p_{1/2}^1            $ &       53581  &          2    &           -10     &         -765    &          2 &      -281  &       45   &        387  &      -376 &     -3903    &      22      \\
$6d_{3/2}^3 6d_{5/2}^6 7s_{1/2}^2 7p_{3/2}^1            $ &       75273  &          1    &            -8     &         -600    &          0 &         7  &       52   &        437  &       -84 &     -1515    &    -126      \\
$6d_{3/2}^3 6d_{5/2}^6 7s_{1/2}^2 8s_{1/2}^1            $ &       85677  &          2    &           -12     &         -915    &         -1 &        25  &       60   &        477  &      -213 &     -2126    &      22      \\
$6d_{3/2}^3 6d_{5/2}^6 7s_{1/2}^2 7d_{3/2}^1            $ &       95546  &          2    &           -10     &         -805    &         -1 &        43  &       60   &        484  &      -119 &     -1760    &       7      \\
\hline
\hline
\end{tabular}
\end{center}
\begin{flushleft}
\noindent (a,c) All-electron Hartree-Fock-Dirac-Breit (HFDB) calculations
              with Fermi and point nuclear charge distributions, accordingly.\\
\noindent (b) All-electron HFDB calculation 
              with the uniform nuclear charge distribution within a sphere. \\
\noindent (d) All-electron HFD calculation 
              with accounting for the Breit interaction
              within PT-1 (HFD+B) and
              with Fermi nuclear model. \\
\noindent (e) All-electron HFD calculation 
              without accounting for the Breit interaction (HFD) and
              with Fermi nuclear model. \\
\noindent (f) GRECP generated in the present work from HFDB calculation 
	      with Fermi nuclear model. \\
\noindent (g) Semi-local RECP generated here
	      from HFDB calculation with Fermi nuclear model
	      on the ionic closed-shell generator-state. \\
\noindent (h) RECP from~\cite{Nash:97} generated from HFD calculation. \\
\noindent (i) PP from M.~Seth {\sl et~al.}
              to be published (P.~Schwerdtfeger,
              private communication, 2003) generated from HFD+B calculation. 
\end{flushleft}
\end{table}

 \begin{table}
\caption{
  Transition Energies (TE) between states averaged over the relativistic 
  configurations of E113 (in cm$^{-1}$). See Table \ref{E112_conf}.
}
\label{E113_conf}
\begin{center}
\begin{tabular}{lrrrrrrrrrr}
\hline
\hline
                                                          &    HFDB         &  HFDB          & HFDB           &   HFDB          &   HFD+B     &   HFD       &           & Ionic     & 21e-RECP    & 21e-PP        \\
                                                          &  (Fermi,        &  (Ball,        & (Fermi,        &   (Point)       &  (Fermi,    &   (Fermi,   & 21e-GRECP & 21e-RECP  & of Nash     & of Seth       \\
                                                          &   A=297)        &  A=297)        & A=284)         &                 &   A=297)    &    A=297)   &           &           & {\it et al.}& {\it et al.}  \\
\hline                                                                                                                                                                                               
 Configuration                                            &  TE             &                                                     \multicolumn{9}{c}{  Absolute errors}                                           \\
\hline                                                                                                                                                                                                            
$6d_{3/2}^4 6d_{5/2}^6 7s_{1/2}^2 7p_{1/2}^1 \rightarrow$ &                 &                &                &                 &             &            &            &           &             &               \\   
$6d_{3/2}^4 6d_{5/2}^6 7s_{1/2}^2 7p_{3/2}^1            $ &          25106  &           0    &         3      &           221   &          -2 &       339  &       -21  &       233 &       275   &        -349   \\
$6d_{3/2}^4 6d_{5/2}^6 7s_{1/2}^2 8s_{1/2}^1            $ &          34981  &           0    &        -2      &          -128   &          -3 &       354  &         5  &       112 &      -205   &        -307   \\
$6d_{3/2}^4 6d_{5/2}^6 7s_{1/2}^2 7d_{3/2}^1            $ &          45172  &           0    &         0      &            -4   &          -3 &       374  &         9  &       200 &       140   &        -275   \\
$6d_{3/2}^4 6d_{5/2}^6 7s_{1/2}^2 6f_{5/2}^1            $ &          50338  &           0    &         0      &           -10   &          -3 &       374  &         9  &       196 &       127   &        -276   \\
$6d_{3/2}^4 6d_{5/2}^6 7s_{1/2}^2 5g_{7/2}^1            $ &          52811  &           0    &         0      &           -10   &          -3 &       374  &         9  &       196 &       127   &        -276   \\
$6d_{3/2}^4 6d_{5/2}^6 7s_{1/2}^2                       $ &          57201  &           0    &         0      &           -10   &          -3 &       374  &         9  &       196 &       127   &        -276   \\
\hline                                                                                                                                                                                                            
$6d_{3/2}^4 6d_{5/2}^6 7s_{1/2}^1 7p_{1/2}^2            $ &          61500  &          -4    &        32      &          2220   &           2 &       -60  &        28  &       610 &      4830   &         148   \\
$6d_{3/2}^4 6d_{5/2}^6 7s_{1/2}^1 7p_{1/2}^1 7p_{3/2}^1 $ &          83184  &          -5    &        36      &          2485   &          -1 &       241  &        -6  &       833 &      5170   &        -172   \\
$6d_{3/2}^4 6d_{5/2}^6 7s_{1/2}^1            7p_{3/2}^2 $ &         112678  &          -6    &        41      &          2843   &          -3 &       612  &       -10  &      1171 &      5717   &        -504   \\
$6d_{3/2}^4 6d_{5/2}^6 7s_{1/2}^1 7p_{1/2}^1            $ &         115758  &          -5    &        34      &          2344   &          -1 &       250  &        -3  &       784 &      5143   &        -105   \\
$6d_{3/2}^4 6d_{5/2}^6 7s_{1/2}^1 7p_{3/2}^1            $ &         149550  &          -5    &        40      &          2739   &          -3 &       654  &        -9  &      1163 &      5784   &        -454   \\
$6d_{3/2}^4 6d_{5/2}^6 7s_{1/2}^1                       $ &         234435  &          -5    &        37      &          2583   &          -4 &       747  &        -2  &      1221 &      6102   &        -336   \\
\hline
\hline                                                                                                                                                                                                            
$6d_{3/2}^4 6d_{5/2}^5 7s_{1/2}^2 7p_{1/2}^2            $ &          47371  &           2    &       -13      &          -864   &           3 &      -739  &       404  &      -597 &     -2349   &         322   \\
$6d_{3/2}^4 6d_{5/2}^5 7s_{1/2}^2 7p_{1/2}^1 7p_{3/2}^1 $ &          74898  &           1    &        -9      &          -606   &           1 &      -391  &       344  &      -378 &     -2055   &         -44   \\
$6d_{3/2}^4 6d_{5/2}^5 7s_{1/2}^2            7p_{3/2}^2 $ &         110406  &           1    &        -4      &          -244   &          -2 &        22  &       310  &       -47 &     -1528   &        -407   \\
$6d_{3/2}^4 6d_{5/2}^5 7s_{1/2}^2 7p_{1/2}^1            $ &         110120  &           2    &       -13      &          -882   &           0 &      -388  &       386  &      -451 &     -2298   &          41   \\
$6d_{3/2}^4 6d_{5/2}^5 7s_{1/2}^2 7p_{3/2}^1            $ &         150102  &           1    &        -7      &          -477   &          -2 &        59  &       344  &       -82 &     -1667   &        -339   \\
$6d_{3/2}^4 6d_{5/2}^5 7s_{1/2}^2                       $ &         239523  &           2    &       -12      &          -807   &          -2 &       144  &       416  &       -39 &     -1617   &        -188   \\
\hline                                                                                                                                                                                                                                  
$6d_{3/2}^3 6d_{5/2}^6 7s_{1/2}^2 7p_{1/2}^2            $ &          78821  &           2    &       -15      &          -983   &           2 &      -375  &       380  &      -649 &     -2230   &        -270   \\
$6d_{3/2}^3 6d_{5/2}^6 7s_{1/2}^2 7p_{1/2}^1 7p_{3/2}^1 $ &         104059  &           1    &       -11      &          -742   &           0 &       -49  &       412  &      -364 &     -1931   &        -544   \\
$6d_{3/2}^3 6d_{5/2}^6 7s_{1/2}^2            7p_{3/2}^2 $ &         137048  &           1    &        -6      &          -403   &          -2 &       341  &       481  &        38 &     -1402   &        -804   \\
$6d_{3/2}^3 6d_{5/2}^6 7s_{1/2}^2 7p_{1/2}^1            $ &         139819  &           2    &       -15      &         -1021   &           0 &       -42  &       447  &      -439 &     -2161   &        -466   \\
$6d_{3/2}^3 6d_{5/2}^6 7s_{1/2}^2 7p_{3/2}^1            $ &         177137  &           1    &        -9      &          -638   &          -3 &       381  &       516  &         9 &     -1523   &        -736   \\
\hline
\hline
\end{tabular}
\end{center}
 \end{table}

 \begin{table}
\caption{
  Transition Energies (TE) between states averaged over the relativistic 
  configurations of E114 (in cm$^{-1}$). See Table \ref{E112_conf}.
}
\label{E114_conf}
\begin{center}
\begin{tabular}{lrrrrrrrrrr}
\hline
\hline
                                                          &   HFDB           &       HFDB    & HFDB      &        HFDB   &  HFD+B     &  HFD       &            & Ionic     & 22e-RECP                   & 22e-PP          \\
                                                          &  (Fermi,         &       (Ball,  & (Fermi,   &       (Point) &  (Fermi,   & (Fermi,    &  22e-GRECP &  22e-RECP & of Nash                    & of Seth         \\
                                                          &   A=298)         &       A=298)  &  A=289)   &               &  A=298)    &  A=298)    &            &           &  {\it et al.}              &  {\it et al.}   \\
\hline                                                                                                                                                                                                                                     
 Configuration                                            & TE               &\multicolumn{9}{c}{  Absolute errors}                                                                                                        \\
\hline                                                                                                                                                                                                                           
$6d_{3/2}^4 6d_{5/2}^6 7s_{1/2}^2 7p_{1/2}^2 \rightarrow$ &                  &               &           &               &            &            &            &           &                            &                 \\ 
$6d_{3/2}^4 6d_{5/2}^6 7s_{1/2}^2 7p_{1/2}^1 7p_{3/2}^1 $ &           29093  &            -1 &         3 &           314 &         -2 &       380  &       -46  &       211 &       449                  &     -457        \\
$6d_{3/2}^4 6d_{5/2}^6 7s_{1/2}^2 7p_{1/2}^1 8s_{1/2}^1 $ &           41211  &             0 &        -1 &          -135 &         -3 &       370  &       -51  &        53 &      -333                  &     -348        \\
$6d_{3/2}^4 6d_{5/2}^6 7s_{1/2}^2 7p_{1/2}^1 8p_{1/2}^1 $ &           48149  &             0 &         0 &            -5 &         -3 &       360  &       -44  &       114 &        11                  &     -318        \\
$6d_{3/2}^4 6d_{5/2}^6 7s_{1/2}^2 7p_{1/2}^1 7d_{3/2}^1 $ &           52230  &             0 &         0 &            16 &         -3 &       387  &       -44  &       149 &       115                  &     -320        \\
$6d_{3/2}^4 6d_{5/2}^6 7s_{1/2}^2 7p_{1/2}^1 6f_{5/2}^1 $ &           57618  &             0 &         0 &             7 &         -3 &       384  &       -43  &       143 &        86                  &     -318        \\
$6d_{3/2}^4 6d_{5/2}^6 7s_{1/2}^2 7p_{1/2}^1 5g_{7/2}^1 $ &           60094  &             0 &         0 &             7 &         -3 &       384  &       -43  &       143 &        86                  &     -317        \\
$6d_{3/2}^4 6d_{5/2}^6 7s_{1/2}^2 7p_{1/2}^1            $ &           64483  &             0 &         0 &             7 &         -3 &       384  &       -43  &       143 &        86                  &     -317        \\
$6d_{3/2}^4 6d_{5/2}^6 7s_{1/2}^2 7p_{3/2}^2            $ &           66669  &            -2 &         6 &           755 &         -5 &       833  &       -52  &       535 &      1209                  &     -926        \\
$6d_{3/2}^4 6d_{5/2}^6 7s_{1/2}^2 7p_{3/2}^1 8s_{1/2}^1 $ &           81879  &            -1 &         3 &           277 &         -5 &       850  &       -64  &       377 &       374                  &     -834        \\
$6d_{3/2}^4 6d_{5/2}^6 7s_{1/2}^2 7p_{3/2}^1            $ &          106776  &            -1 &         4 &           497 &         -5 &       872  &       -53  &       503 &       969                  &     -808        \\
$6d_{3/2}^4 6d_{5/2}^6 7s_{1/2}^2 8s_{1/2}^2            $ &          108893  &             1 &        -3 &          -361 &         -6 &       883  &       -82  &       174 &      -725                  &     -718        \\
$6d_{3/2}^4 6d_{5/2}^6 7s_{1/2}^2 8s_{1/2}^1            $ &          136567  &             0 &        -1 &          -207 &         -6 &       907  &       -72  &       285 &      -241                  &     -680        \\
$6d_{3/2}^4 6d_{5/2}^6 7s_{1/2}^2                       $ &          197486  &             0 &         1 &           128 &         -6 &       961  &       -45  &       547 &       853                  &     -584        \\
\hline                                                                                                                                                                                                                             
$6d_{3/2}^4 6d_{5/2}^6 7s_{1/2}^1 7p_{1/2}^2 7p_{3/2}^1 $ &          102896  &            -6 &        24 &          3110 &          0 &       256  &        96  &       929 &      6650                  &     -327        \\
$6d_{3/2}^4 6d_{5/2}^6 7s_{1/2}^1 7p_{1/2}^2 8s_{1/2}^1 $ &          115405  &            -5 &        21 &          2745 &         -1 &       224  &        59  &       754 &      5987                  &     -243        \\
$6d_{3/2}^4 6d_{5/2}^6 7s_{1/2}^1 7p_{1/2}^2            $ &          138842  &            -6 &        23 &          2905 &         -1 &       233  &        73  &       848 &      6439                  &     -206        \\
\hline
\hline                                                                                                                                                                                                                       
$6d_{3/2}^4 6d_{5/2}^5 7s_{1/2}^2 7p_{1/2}^2 7p_{3/2}^1 $ &           97736  &             2 &        -6 &          -771 &          1 &      -506  &       472  &      -631 &     -3156                  &      -28        \\
$6d_{3/2}^4 6d_{5/2}^5 7s_{1/2}^2 7p_{1/2}^2 8s_{1/2}^1 $ &          112486  &             3 &       -10 &         -1277 &          1 &      -543  &       473  &      -830 &     -4091                  &       83        \\
$6d_{3/2}^4 6d_{5/2}^5 7s_{1/2}^2 7p_{1/2}^2            $ &          136356  &             2 &        -9 &         -1129 &          1 &      -534  &       487  &      -732 &     -3647                  &      122        \\
\hline                                                                                                                                                                                                                                  
$6d_{3/2}^3 6d_{5/2}^6 7s_{1/2}^2 7p_{1/2}^2 7p_{3/2}^1 $ &          133837  &             2 &        -7 &          -904 &          0 &       -91  &       391  &      -746 &     -2821                  &     -675        \\
$6d_{3/2}^3 6d_{5/2}^6 7s_{1/2}^2 7p_{1/2}^2 8s_{1/2}^1 $ &          149162  &             3 &       -11 &         -1415 &          0 &      -126  &       380  &      -952 &     -3758                  &     -579        \\
$6d_{3/2}^3 6d_{5/2}^6 7s_{1/2}^2 7p_{1/2}^2            $ &          173108  &             3 &       -10 &         -1265 &          0 &      -117  &       391  &      -855 &     -3309                  &     -541        \\
\hline
\hline
\end{tabular}
\end{center}
 \end{table}

\begin{table}
\caption{
Transition Energies (TE) between states averaged over the nonrelativistic 
configurations of uranium (in cm$^{-1}$). See footnotes in Table \ref{E112_conf}.
}
\label{U_conf}
\begin{center}
\begin{tabular}{lrrrrr}
\hline
\hline
                              &  HFDB   &  HFDB   & HFD+B   &  HFD    &           \\
                              & (Fermi, & (Point) & (Fermi, & (Fermi, & 32e-GRECP \\
                              &  A=238) &         & A=238)  &  A=238) &           \\
\hline                                                                                                                                                                                                                                     
 Configuration                & TE      &\multicolumn{4}{c}{Absolute errors}      \\
\hline                                                                                                                                                                                                                           
 $5f^3 7s^2 6d^1 \rightarrow$ &         &         &         &         &          \\
 $5f^3 7s^2 7p^1            $ &    7516 &    -40  &       0 &     -93 &        5 \\   
 $5f^3 7s^2                 $ &   36289 &    -68  &       0 &     -62 &        9 \\   
 $5f^3 7s^1 6d^2            $ &   13124 &     97  &       0 &      78 &       -7 \\   
 $5f^3 7s^1 6d^1 7p^1       $ &   17200 &     75  &       0 &      14 &       -1 \\   
 $5f^3 7s^1 6d^1            $ &   42328 &     63  &       0 &      44 &        0 \\   
 $5f^3 6d^2                 $ &   54576 &    177  &       0 &     138 &       -6 \\   
 $5f^3 7s^2 6d^1 \rightarrow$ &         &         &         &         &          \\
 $5f^4 7s^2                 $ &   15780 &     76  &       2 &     627 &     -363 \\   
 $5f^4 7s^2      \rightarrow$ &         &         &         &         &          \\
 $5f^4 7s^1 6d^1            $ &   15010 &     78  &       0 &      43 &        3 \\   
 $5f^4 7s^1 7p^1            $ &   14932 &     62  &       0 &      21 &       -3 \\   
 $5f^4 7s^1                 $ &   38813 &     50  &      -1 &      50 &       -3 \\   
 $5f^4 6d^2                 $ &   33792 &    147  &       1 &      82 &        6 \\   
 $5f^4 6d^1 7p^1            $ &   32115 &    146  &       0 &      79 &        2 \\   
 $5f^4 6d^1                 $ &   53379 &    148  &       0 &     108 &        1 \\   
 $5f^3 7s^2 6d^1 \rightarrow$ &         &         &         &         &          \\
 $5f^2 7s^2 6d^2            $ &    4640 &    -85  &      -1 &    -779 &      362 \\   
 $5f^2 7s^2 6d^2 \rightarrow$ &         &         &         &         &          \\
 $5f^2 7s^2 6d^1 7p^1       $ &   12809 &    -44  &       0 &    -118 &       11 \\   
 $5f^2 7s^2 6d^1            $ &   42793 &    -71  &       0 &     -83 &       15 \\   
 $5f^2 7s^1 6d^3            $ &   10480 &    113  &       0 &     104 &      -12 \\   
 $5f^2 7s^1 6d^2 7p^1       $ &   19217 &     87  &       0 &      15 &       -1 \\   
 $5f^2 7s^1 6d^2            $ &   45352 &     75  &       0 &      50 &        0 \\   
 $5f^2 6d^3                 $ &   54611 &    204  &       0 &     168 &      -12 \\   
 $5f^3 7s^2 6d^1 \rightarrow$ &         &         &         &         &          \\
 $5f^1 7s^2 6d^3            $ &   31450 &   -176  &      -2 &   -1673 &      680 \\   
 $5f^1 7s^2 6d^3 \rightarrow$ &         &         &         &         &          \\
 $5f^1 7s^2 6d^2 7p^1       $ &   18326 &    -48  &       0 &    -137 &       11 \\   
 $5f^1 7s^2 6d^2            $ &   49329 &    -75  &       0 &     -96 &       16 \\   
 $5f^1 7s^1 6d^4            $ &    7331 &    127  &       0 &     124 &      -15 \\   
 $5f^1 7s^1 6d^3 7p^1       $ &   21038 &     98  &       0 &      18 &       -1 \\   
 $5f^1 7s^1 6d^3            $ &   48001 &     87  &       0 &      57 &        0 \\   
 $5f^1 6d^4                 $ &   53806 &    230  &       0 &     196 &      -15 \\   
 $5f^3 7s^2 6d^1 \rightarrow$ &         &         &         &         &          \\
 $5f^5                      $ &   99459 &    252  &       4 &    1126 &     -671 \\
\hline
\hline
\end{tabular}
\end{center}
\end{table}

\begin{table}
\caption{
Transition Energies (TE) between states averaged over the nonrelativistic 
configurations of plutonium (in cm$^{-1}$). See
 footnotes in Table \ref{E112_conf}.
}
\label{Pu_conf}
\begin{center}
\begin{tabular}{lrrrrr}
\hline
\hline
                              &  HFDB   &  HFDB   & HFD+B   &  HFD    &           \\
                              & (Fermi, & (Point) & (Fermi, & (Fermi, & 34e-GRECP \\
                              &  A=244) &         & A=244)  &  A=244) &           \\
\hline                                                                                                                                                                                                                                     
 Configuration                & TE      &\multicolumn{4}{c}{Absolute errors}      \\
\hline                                                                                                                                                                                                                           
 $5f^6 7s^2      \rightarrow$ &         &         &         &         &           \\          
 $5f^6 7s^1 6d^1            $ &   17164 &      96 &       0 &      53 &        -2 \\  
 $5f^6 7s^1      7p^1       $ &   15678 &      76 &       0 &      19 &        -1 \\     
 $5f^6 7s^1                 $ &   39853 &      61 &       0 &      47 &        -1 \\     
 $5f^6      6d^1            $ &   56794 &     183 &       0 &     114 &        -2 \\     
 $5f^6           7p^1       $ &   66677 &     172 &      -1 &      71 &         1 \\     
 $5f^6 7s^2      \rightarrow$ &         &         &         &         &           \\
 $5f^7 7s^1                 $ &   43691 &     159 &       4 &     504 &      -377 \\     
 $5f^7 7s^1      \rightarrow$ &         &         &         &         &           \\          
 $5f^7      6d^1            $ &   19877 &      67 &       0 &      54 &        -1 \\     
 $5f^7           7p^1       $ &   14816 &      68 &      -1 &      62 &        -6 \\     
 $5f^7                      $ &   34957 &      70 &      -1 &      96 &        -9 \\     
 $5f^6 7s^2      \rightarrow$ &         &         &         &         &           \\
 $5f^5 7s^2 6d^1            $ &   -3099 &    -103 &      -2 &    -704 &       414 \\     
 $5f^5 7s^2 6d^1 \rightarrow$ &         &         &         &         &           \\          
 $5f^5 7s^2      7p^1       $ &    6743 &     -50 &       0 &     -93 &        10 \\     
 $5f^5 7s^1 6d^2            $ &   15044 &     120 &       0 &      82 &       -10 \\     
 $5f^5 7s^1 6d^1 7p^1       $ &   18246 &      94 &       0 &      17 &         0 \\     
 $5f^5 7s^2                 $ &   35910 &     -84 &       0 &     -61 &        14 \\     
 $5f^5 7s^1 6d^1            $ &   43764 &      80 &       0 &      48 &         1 \\     
 $5f^6 7s^2      \rightarrow$ &         &         &         &         &           \\
 $5f^4 7s^2 6d^2            $ &   17425 &    -213 &      -2 &   -1545 &       807 \\     
 $5f^4 7s^2 6d^2 \rightarrow$ &         &         &         &         &           \\          
 $5f^4 7s^2 6d^1 7p^1       $ &   12434 &     -55 &       0 &    -116 &        16 \\     
 $5f^4 7s^1 6d^3            $ &   12221 &     141 &       0 &     105 &       -16 \\     
 $5f^4 7s^1 6d^2 7p^1       $ &   20405 &     109 &       0 &      18 &        -1 \\     
 $5f^4 7s^2 6d^1            $ &   42841 &     -88 &       0 &     -77 &        19 \\     
 $5f^4 7s^1 6d^2            $ &   46949 &      95 &       0 &      55 &         0 \\     
 $5f^6 7s^2      \rightarrow$ &         &         &         &         &           \\
 $5f^3 7s^2 6d^3            $ &   62648 &    -328 &      -3 &   -2496 &      1136 \\     
 $5f^3 7s^2 6d^3 \rightarrow$ &         &         &         &         &           \\          
 $5f^3 7s^1 6d^4            $ &    8926 &     159 &       0 &     124 &       -20 \\     
 $5f^3 7s^2 6d^2 7p^1       $ &   18247 &     -59 &       0 &    -133 &        18 \\     
 $5f^3 7s^1 6d^3 7p^1       $ &   22323 &     123 &       0 &      21 &        -1 \\     
 $5f^3      6d^5            $ &   24140 &     295 &       0 &     231 &       -36 \\     
 $5f^3 7s^2 6d^2            $ &   49677 &     -92 &       0 &     -89 &        22 \\     
 $5f^3 7s^1 6d^3            $ &   49694 &     109 &       0 &      63 &         0 \\     
\hline
\hline
\end{tabular}
\end{center}
\end{table}

 \begin{table}
\caption{
  Transition Energies (TE) between terms of E112 (in cm$^{-1}$). See
  footnotes in Table \ref{E112_conf}.
}
\label{E112_term}
\vspace{5mm}
\begin{center}
\begin{tabular}{lrrrrrrrr}
\hline
\hline
                     &  HFDB   & HFD+B   &  HFD    & 52e-      & 20e-      &Ionic\,20e-& 20e-RECP    & 20e-PP       \\
                     & (Fermi, & (Fermi, & (Fermi, & GRECP     & GRECP     &     RECP & of Nash      & of Seth      \\
                     &  A=296) & A=296)  &  A=296) &           &           &          & {\it et al.} & {\it et al.} \\
\hline                                                                                                                                                                                                                                     
 Configuration, term & TE      &\multicolumn{7}{c}{Absolute errors}                                                 \\
\hline                                                                                                                                                                                                                           
 $6d_{3/2}^4 6d_{5/2}^6 7s_{1/2}^1 7p_{1/2}^1$ J=0 $\rightarrow$ & & & & & &          &              &              \\
 J=1                 &    9468 &       0 &      54 &   9       &        42 &       59 &          288 &           27 \\ 
 $6d_{3/2}^4 6d_{5/2}^5 7s_{1/2}^2 7p_{1/2}^1$ J=2 $\rightarrow$ & & & & & &          &              &              \\
 J=3                 &    1958 &       0 &      25 &   6       &        16 &       43 &          165 &           11 \\ 
 $6d_{3/2}^3 6d_{5/2}^6 7s_{1/2}^2 7p_{1/2}^1$ J=1 $\rightarrow$ & & & & & &          &              &              \\
 J=2                 &   -8145 &       1 &     -92 &   3       &       172 &       40 &         -558 &          100 \\ 
 $6d_{3/2}^4 6d_{5/2}^5 7s_{1/2}^2 7p_{3/2}^1$ J=1 $\rightarrow$ & & & & & &          &              &              \\
 J=2                 &   -1919 &       0 &     -24 &  -5       &        34 &       14 &          -42 &           16 \\ 
 J=3                 &      39 &       0 &     -17 &   0       &        74 &       56 &            4 &           78 \\ 
 J=4                 &   -3166 &       0 &     -27 &  -6       &         9 &      -13 &          -69 &          -23 \\ 
\hline
\hline
\end{tabular}
\end{center}
 \end{table}

\begin{table}
\caption{
Transition Energies (TE) between terms of uranium (in cm$^{-1}$). See
 footnotes in Table \ref{E112_conf}.
}
\label{U_term}
\begin{center}
\begin{tabular}{lrrrrrrr}
\hline
\hline
                     &  HFDB   & HFD+B   &  HFD    & 32e-      & 32e-      & 24e-SfC & 24e-SfC   \\
                     & (Fermi, & (Fermi, & (Fermi, & GRECP     & GRECP     &   GRECP &     GRECP \\
                     &  A=238) & A=238)  &  A=238) &           & TS-corr.\ &         & TS-corr.\ \\
		     &         &         &         &           & (a)       & (b)     &   (a,b)   \\
\hline                                                                                                                                                                                                                                     
 Configuration, term & TE      &\multicolumn{6}{c}{Absolute errors}                              \\
\hline                                                                                                                                                                                                                           
\multicolumn{2}{l}{
 $5f_{5/2}^3 6d_{3/2}^1 7s_{1/2}^2$ J=0 $\rightarrow$ } & & &  &           &         &           \\
 J=1                 &   18576 &       0 &      74 &       137 &       -15 &      67 &      -102 \\ 
 J=2                 &    9710 &       0 &      22 &       140 &       -12 &     117 &       -53 \\ 
 J=3                 &    7749 &       0 &      66 &       -57 &        -9 &    -103 &       -49 \\ 
 J=4                 &    6691 &       0 &      69 &       -77 &        -5 &    -121 &       -40 \\ 
 J=5                 &   -8005 &       0 &      83 &      -439 &         8 &    -470 &        31 \\ 
 J=6                 &  -10767 &       0 &      69 &      -416 &        31 &    -431 &        69 \\ 
\multicolumn{2}{l}{
 $5f_{5/2}^3 5f_{7/2}^1 7s_{1/2}^2$ J=1 $\rightarrow$ } & & &  &           &         &           \\                                                                                                                
 J=2                 &    4399 &       0 &      -5 &       159 &       -35 &     165 &       -51 \\ 
 J=3                 &    2840 &       0 &       4 &       109 &       -23 &     113 &       -33 \\ 
 J=4                 &    3468 &       0 &      11 &       134 &       -29 &     139 &       -42 \\ 
 J=5                 &    2785 &       0 &      22 &       117 &       -24 &     121 &       -36 \\ 
 J=6                 &    4606 &       1 &      29 &       181 &       -42 &     188 &       -62 \\  
 J=7                 &   -6030 &       1 &      78 &      -176 &        12 &    -186 &        26 \\ 
 J=8                 &   -5542 &       1 &      90 &      -149 &         6 &    -158 &        17 \\ 
\multicolumn{2}{l}{
 $5f_{5/2}^2 6d_{3/2}^2 7s_{1/2}^2$ J=0 $\rightarrow$ } & & &  &           &         &           \\ 
 J=1                 &  -19109 &       0 &      23 &      -432 &       -61 &    -426 &        -8 \\ 
 J=2                 &  -15310 &       0 &       1 &      -304 &       -45 &    -288 &         5 \\ 
 J=3                 &  -23656 &       0 &      41 &      -598 &       -77 &    -593 &        -8 \\ 
 J=4                 &  -26013 &       0 &      21 &      -638 &       -69 &    -618 &        23 \\ 
 J=5                 &  -32544 &       0 &      36 &      -754 &       -86 &    -732 &        21 \\ 
 J=6                 &  -39562 &       0 &      -2 &      -724 &       -57 &    -671 &        82 \\ 
\hline
\hline
\end{tabular}
\end{center}

\begin{flushleft}
\noindent (a) Term-Splitting (TS) correction generated in the present work 
              from HFDB calculation with Fermi nuclear charge distribution.\\
\noindent (b) Self-Consistent Generalized Relativistic Effective Core 
              Potential (SfC~GRECP) generated in \cite{Petrov:04b} 
	      from HFDB calculation with Fermi nuclear charge distribution. 
\end{flushleft}
\end{table}

  \begin{table}
\caption{
Matrix Elements (ME) of $<r^2>$ for some spinors from states averaged over 
the relativistic configurations of E112 (in a.u.).  See
 footnotes in Table \ref{E112_conf}.
}
\label{E112_r2}
\begin{center}
\begin{tabular}{lrrrrrrr}
\hline
\hline
                       &  HFDB   &  HFD    & 52e-      & 20e-      &Ionic\,20e-& 20e-RECP    & 20e-PP       \\
                       & (Fermi, & (Fermi, & GRECP     & GRECP     &     RECP & of Nash      & of Seth      \\
                       &  A=296) &  A=296) &           &           &          & {\it et al.} & {\it et al.} \\
\hline                                                                                                                                                                                                                                     
 Config.,
                spinor & ME      &\multicolumn{6}{c}{Absolute errors}                                       \\
\hline                                                                                                                                                                                                                           
 $6d_{3/2}^4 6d_{5/2}^6 7s_{1/2}^2$ & &    &           &           &          &              &              \\
 $6d_{3/2}$            &   3.150 &  -0.005 &  0.001    &  0.024    &    0.030 &        0.066 &        0.072 \\
 $6d_{5/2}$            &   3.781 &   0.002 &  0.001    &  0.024    &    0.032 &        0.074 &        0.057 \\
 $7s_{1/2}$            &   7.157 &  -0.023 &  0.000    &  0.005    &   -0.099 &       -0.425 &        0.024 \\
\multicolumn{2}{l}{
 $6d_{3/2}^4 6d_{5/2}^6 7s_{1/2}^1 7p_{1/2}^1$   } & & &           &          &              &              \\                                          
 $6d_{3/2}$            &   3.144 &  -0.004 &  0.001    &  0.023    &    0.028 &        0.064 &        0.071 \\
 $6d_{5/2}$            &   3.648 &   0.002 &  0.001    &  0.024    &    0.031 &        0.069 &        0.057 \\
 $7s_{1/2}$            &   6.898 &  -0.020 &  0.000    &  0.002    &   -0.097 &       -0.394 &        0.022 \\
 $7p_{1/2}$            &  13.023 &  -0.116 & -0.001    &  0.005    &   -0.131 &       -0.841 &        0.055 \\
\multicolumn{2}{l}{
 $6d_{3/2}^4 6d_{5/2}^5 7s_{1/2}^2 7p_{1/2}^1$   } & & &           &          &              &              \\                                          
 $6d_{3/2}$            &   3.057 &  -0.004 &  0.001    &  0.023    &    0.030 &        0.071 &        0.071 \\
 $6d_{5/2}$            &   3.522 &   0.002 &  0.001    &  0.025    &    0.035 &        0.080 &        0.059 \\
 $7s_{1/2}$            &   6.739 &  -0.019 &  0.000    &  0.001    &   -0.092 &       -0.361 &        0.025 \\
 $7p_{1/2}$            &  11.259 &  -0.087 & -0.002    & -0.001    &   -0.105 &       -0.597 &        0.049 \\
\hline
\hline
\end{tabular}
\end{center}
  \end{table}

\begin{table}
\caption{Radial integrals
$2\int^{\infty}_{R_{n}}\,{\mid}\,
f_{n_vlj}(r) [f_{n_vlj}(r) - \widetilde{\varphi}_{n_vlj}(r)]\,{\mid}\,dr$
for valence spinors from states averaged over 
the relativistic configurations of E112 (in a.u.)
where $f_{n_vlj}$ is the large component of the Dirac spinor from HFDB 
calculation with the Fermi nuclear charge distribution for $A=296$,
$\widetilde{\varphi}_{n_vlj}$ is the radial part of the corresponding 
pseudospinor (or the large component of the Dirac spinor),
$R_{n}$ is the radius of the last node for the spinor. 
See footnotes in Table \ref{E112_conf}.
}
\label{E112_int}
\begin{center}
\begin{tabular}{lrrrrrr}
\hline
\hline
                       &  HFD    & 52e-      &  20e-     &Ionic\,20e-& 20e-RECP    & 20e-PP       \\
                       & (Fermi, &GRECP      & GRECP     &     RECP & of Nash      & of Seth      \\
                       &  A=296) &           &           &          & {\it et al.} & {\it et al.} \\
\hline                                                                                                                                                                                                                                     
 Configuration, spinor &\multicolumn{6}{c}{Integrals}                                             \\
\hline                                                                                                                                                                                                                           
 $6d_{3/2}^4 6d_{5/2}^6 7s_{1/2}^2$ & &      &           &          &              &              \\
 $7s_{1/2}$            &  0.0037 &  0.0000   &  0.0006   &   0.0131 &       0.0590 &       0.0024 \\
 $6d_{3/2}^4 6d_{5/2}^6 7s_{1/2}^1 7p_{1/2}^1$ & & &     &          &              &              \\                                          
 $7s_{1/2}$            &  0.0036 &  0.0001   &  0.0002   &   0.0137 &       0.0581 &       0.0022 \\
 $7p_{1/2}$            &  0.0087 &  0.0001   &  0.0004   &   0.0091 &       0.0610 &       0.0047 \\
 $6d_{3/2}^4 6d_{5/2}^5 7s_{1/2}^2 7p_{1/2}^1$ & & &     &          &              &              \\                                          
 $7s_{1/2}$            &  0.0034 &  0.0001   &  0.0003   &   0.0134 &       0.0546 &       0.0026 \\
 $7p_{1/2}$            &  0.0079 &  0.0002   &  0.0002   &   0.0088 &       0.0514 &       0.0050 \\
\hline
\hline
\end{tabular}
\end{center}
\end{table}

\begin{table}
\caption{
One-electron energies, $\varepsilon$, for some spinors from states averaged 
over the relativistic configurations of
 E112 (in a.u.).  See footnotes in Table \ref{E112_conf}.
}
\label{E112_orb}
\begin{center}
\begin{tabular}{lrrrrrrr}
\hline
\hline
                       &  HFDB         &  HFD    &  52e-     &   20e-    &Ionic\,20e-& 20e-RECP    & 20e-PP       \\
                       & (Fermi,       & (Fermi, & GRECP     &  GRECP    & RECP     & of Nash      & of Seth      \\
                       &  A=296)       &  A=296) &           &           &          & {\it et al.} & {\it et al.} \\
\hline                                                                                                                                                                                                                                     
 Config.,
 spinor & $\varepsilon$ &\multicolumn{6}{c}{Absolute errors}                                       \\
\hline                                                                                                                                                                                                                           
 $6d_{3/2}^4 6d_{5/2}^6 7s_{1/2}^2$ & &          &           &           &          &              &              \\
 $6d_{3/2}$            &       0.5624 &   0.0003 &  0.0000   &    0.0001 &  -0.0026 &      -0.0112 &      -0.0037 \\
 $6d_{5/2}$            &       0.4432 &  -0.0011 &  0.0000   &   -0.0001 &  -0.0026 &      -0.0095 &      -0.0009 \\
 $7s_{1/2}$            &       0.4497 &   0.0014 &  0.0000   &   -0.0003 &   0.0042 &       0.0272 &       0.0001 \\
\multicolumn{2}{l}{
 $6d_{3/2}^4 6d_{5/2}^6 7s_{1/2}^1 7p_{1/2}^1$ } &           &           &          &              &              \\                                          
 $6d_{3/2}$            &       0.6148 &  -0.0001 &  0.0000   &   -0.0001 &  -0.0025 &      -0.0118 &      -0.0039 \\
 $6d_{5/2}$            &       0.4870 &  -0.0017 &  0.0000   &    0.0000 &  -0.0024 &      -0.0108 &      -0.0009 \\
 $7s_{1/2}$            &       0.5217 &   0.0011 &  0.0000   &    0.0000 &   0.0048 &       0.0278 &       0.0004 \\
 $7p_{1/2}$            &       0.2248 &   0.0015 &  0.0000   &   -0.0001 &   0.0012 &       0.0114 &      -0.0004 \\
\multicolumn{2}{l}{
 $6d_{3/2}^4 6d_{5/2}^5 7s_{1/2}^2 7p_{1/2}^1$ } &           &           &          &              &              \\
 $6d_{3/2}$            &       0.6663 &  -0.0002 &  0.0000   &   -0.0002 &  -0.0042 &      -0.0173 &      -0.0043 \\
 $6d_{5/2}$            &       0.5314 &  -0.0018 &  0.0000   &   -0.0001 &  -0.0040 &      -0.0159 &      -0.0014 \\
 $7s_{1/2}$            &       0.5253 &   0.0010 &  0.0000   &    0.0000 &   0.0044 &       0.0258 &       0.0001 \\
 $7p_{1/2}$            &       0.2653 &   0.0017 &  0.0000   &    0.0000 &   0.0011 &       0.0110 &      -0.0005 \\
\hline
\hline
\end{tabular}
\end{center}
\end{table}

\end{document}